\documentclass[12pt]{article}

\usepackage{newtxtext,newtxmath}

\usepackage{graphicx}

\usepackage[letterpaper,margin=1in]{geometry}

\renewenvironment{abstract}
	{\quotation}
	{\endquotation}

\date{}

\makeatletter
\renewcommand{\fnum@figure}{\textbf{Figure \thefigure}}
\renewcommand{\fnum@table}{\textbf{Table \thetable}}
\makeatother

\usepackage{scicite}

\usepackage{url}

\usepackage[abbreviations]{glossaries-extra} 
\glssetcategoryattribute{abbreviation}{nohyper}{true}
\newabbreviation{cuore}{CUORE}{Cryogenic Underground Observatory for Rare Events}

\newcommand{\ndbd}{\ensuremath{0\nu \beta\beta}}
\newcommand{\ndbdbold}{\ensuremath{\boldsymbol{0\nu \beta\beta}}}
\newcommand{\dbd}{\ensuremath{2\nu \beta\beta}}
\newcommand{\qbb}{\ensuremath{Q_{\beta\beta}}}

\newcommand{\mbb}{\ensuremath{m_{\beta\beta}}}
\newcommand{\mbbbold}{\ensuremath{\boldsymbol{m_{\beta\beta}}}}

\usepackage{subcaption} 
\usepackage{float} 

\usepackage{booktabs} 

\def\scititle{
	Constraints on Lepton Number Violation with the\\ 2 tonne$\cdot$yr CUORE Dataset
}
\title{\bfseries \boldmath \scititle}

\author{
        \ \qquad CUORE~Collaboration${^\ast}\dagger$\ \ \qquad \and
	\small$^\ast$Corresponding author: cuore-spokesperson@lngs.infn.it\and
        \small$\dagger$CUORE Collaboration authors and affiliations are listed in the supplementary materials.
}

\begin{document} 

\maketitle

\begin{abstract} \bfseries \boldmath
Matter-antimatter asymmetry underlines the incompleteness of the current understanding of particle physics. 
Neutrinoless double-beta decay (\ndbdbold) may help explain this asymmetry, while unveiling the Majorana nature of the neutrino. The CUORE experiment searches for \ndbdbold\ of $^{130}$Te using a tonne-scale cryogenic calorimeter operated at milli-kelvin temperatures. We report no evidence of \ndbdbold\ and place a lower limit on the half-life of \ensuremath{\boldsymbol{T_{1/2}}} $>$ 3.5 $\times$ 10$^{25}$~years (90\% C.I.) with over 2~tonne$\cdot$year TeO$_2$ exposure. 
The tools and techniques developed for this result and the 5 year stable operation of nearly 1000 detectors demonstrate crucial infrastructure for a future-generation experiment capable of searching for \ndbdbold\ across multiple isotopes.
\end{abstract}

\noindent
The abundance of baryonic matter in the observable universe contrasts with the leading theories of particle physics~\cite{ParticleDataGroup:2022pth}. Cosmological models assume that matter and antimatter should have been created in equal amounts. When they come into contact with each other, they interact and leave behind only energy. As a result, the universe should be composed of either equal amounts of matter and antimatter or none at all. Observations, however, contradict this hypothesis, indicating a fundamental asymmetry in the properties of matter and antimatter~\cite{Dine:2003ax}. Neutrinos, the least understood of the fundamental particles, may help resolve this impasse if they turn out to be Majorana fermions---indistinguishable from their own antiparticles~\cite{Majorana:1937vz}. 

Double-beta decay (usually labeled \dbd)~\cite{GoeppertMayer:1935qp} is a rare process in which two neutrons in a nucleus, with mass $A$ and atomic number $Z$, convert into two protons with the emission of two electrons and two antineutrinos: $(A, Z) \longrightarrow (A, Z+2)+2e^{-}+2\bar{\nu}_e$. If the neutrino is its own antiparticle, the process could leave matter as the only by-product: $(A, Z) \longrightarrow (A, Z+2)+2e^{-}$. This version of the decay, so-called neutrinoless double-beta decay~(labeled \ndbd)~\cite{Furry:1939qr}, would imply lepton number is not a conserved quantity in nature. As a lepton number violating process, discovery of \ndbd\ would lend support for the theory of leptogenesis in the early universe, which, acting as a potential source of baryogenesis, offers an explanation for the observed matter-antimatter asymmetry~\cite{Sakharov:1967dj}. Moreover, the absence of neutrinos from the decay products would not only confirm the Majorana nature of the neutrino, but also help constrain the mass of this elusive particle. Currently, only \dbd\ has been observed with half-life measurements in the range of $T_{1/2}^{2\nu}\sim 10^{19} - 10^{22}$~yr~\cite{ParticleDataGroup:2022pth,Augier:2023prl,CUPID:2023wyy,CUORE:2020bok,CUORE:2020bok_erratum}. 

Neutrinoless double-beta decay would potentially occur in the same isotopes that exhibit \dbd~\cite{Dolinski:Review}. Experiments searching for \ndbd\ are designed to measure a monoenergetic peak at the Q-value of the decay (\qbb) which would indicate the emission of only two electrons that fully deposit their energies in the detector. Experimental sensitivity is maximized by increasing their measurement time and the mass of the $\beta\beta$ source under observation. When the background---i.e., energy measurements derived from unrelated phenomena---near \qbb\ is non-negligible, experiments must both minimize the background and optimize the detector energy resolution to distinguish the signal peak~\cite{Agostini:2022zub}.

The use of cryogenic calorimeters for \ndbd\ searches was first proposed by Fiorini and Niinikoski in 1984~\cite{Fiorini:1983yj}. In the following decades, the detection technique demonstrated excellent energy resolution ($\sim$0.1\% in the MeV range), radiopurity control, and scalability to large masses ($\sim$1~tonne)~\cite{Brofferio:2018lys,CUORE:2021ctv}. In this application, the calorimeter components consist of an absorber material primarily composed of a $\beta\beta$ source, a sensor to measure the temperature of the absorber, and a weak thermal coupling of the absorber to a heat sink. An example of a cryogenic calorimeter is shown in Fig.~\ref{fig:crystal}. When energy is released in the absorber, its temperature rise is proportional to the energy deposition and inversely proportional to its heat capacity. Because the heat sink serves to restore the detector to a base temperature, over time the energy deposition is represented by a pulse in the sensor readout. To have a measurable temperature rise, the absorber material is kept at very low temperatures (T$\sim$10~mK), where the heat capacity (C) is reduced. Diamagnetic and dielectric materials are preferred because their lattice-specific heat capacity, which follows the Debye law C(T)~$\propto$~T$^3$, is the only contribution, but the choice of the material is otherwise flexible. The flexibility in the choice of the $\beta\beta$ emitter (i.e., $^{48}$Ca, $^{76}$Ge, $^{82}$Se, $^{100}$Mo, $^{116}$Cd, and $^{130}$Te) is a key strength of the cryogenic calorimeter technique. So far, $^{130}$Te~\cite{CUORE:2021mvw}, $^{82}$Se~\cite{CUPID:2022puj}, and $^{100}$Mo~\cite{Augier:2022fr,Agrawal:2025} have been used by cryogenic experiments in high sensitivity searches for \ndbd, leading to some of the most stringent half-life limits on this process. If \ndbd\ is observed, measuring the process across multiple isotopes would be crucial to understanding the new physics involved~\cite{Cremonesi:2013vla}. This possibility continues to propel efforts in 
crystal growing~\cite{Armengaud:2017cry}
and background reduction methods~\cite{ARMATOL2024169936,PhysRevApplied.20.064017} that will allow this technique to meet the sensitivity goals of beyond the next-generation experiments. Thus, cryogenic calorimeters offer a wide pathway toward this long-term objective~\cite{Giuliani:2017tqo}.

\subsection*{The CUORE experiment}

The CUORE (Cryogenic Underground Observatory for Rare Events) experiment~\cite{CUORE:2021mvw} primarily searches for \ndbd\ of $^{130}$Te at the Laboratori Nazionali del Gran Sasso (LNGS) in Italy. The CUORE detector consists of a close-packed array of 988 5\,$\times$\,5\,$\times$\,5~cm$^{3}$ TeO$_2$ crystal calorimeters with an active mass of 742~kg (206~kg $^{130}$Te). The crystals are mounted in copper frames that form 19 towers. The towers are hosted in the CUORE cryostat, a custom multistage cryogen-free dilution refrigerator that cools the detector to $\sim$10~mK and maintains it in a stable, low-noise environment~\cite{Alduino:2019xia}. 
An illustration of the CUORE crystal arrangement is shown in Fig.~\ref{fig:CUORE_3d_exposure}. 

CUORE chose $^{130}$Te as the $\beta\beta$ emitter because of its relatively high natural isotopic abundance of $\sim$34\%, which in turn allowed us to use tellurium with natural isotopic composition. This enables CUORE to search for \ndbd\ and other rare processes in multiple isotopes of tellurium~\cite{CUORE:2020bok,CUORE:2020bok_erratum,CUORE:2021xns,CUORE120Te,CUORE128Te}. The $\beta\beta$ emitter is also favorable because of its relatively high \qbb\ of $\sim$2527.5~keV~\cite{Rahaman:2011zz}, 
which is above most environmental gamma-ray backgrounds. CUORE is further protected by several means from backgrounds that could obscure the \ndbd\ signal. The Gran Sasso mountains above LNGS provide natural shielding from cosmic rays, decreasing the muon flux by $\sim$6 orders of magnitude compared to sea level~\cite{Borexino:2012wej}. Multiple shields located outside and inside the cryostat provide nearly 4$\pi$ coverage against environmental gamma background and neutrons~\cite{Wulandari:2003cr,Alduino:2019xia} with the inner cryostat shield mostly comprised of $^{210}$Pb-depleted ancient 
lead recovered from a Roman shipwreck~\cite{Pattavina:2019pxw}. We addressed the largest source of background---i.e., degraded alpha particles originating from the trace radioactive contamination of the copper frame structures~\cite{CUORE:2024fak}---using  effective cleaning procedures and strict radiopurity controls on the crystals and nearby structures~\cite{Alessandria:2012zp,Buccheri:2014bma,Benato:2017kdf}. 

\begin{figure}[H]
    \centering
    \includegraphics[width=0.6\textwidth]{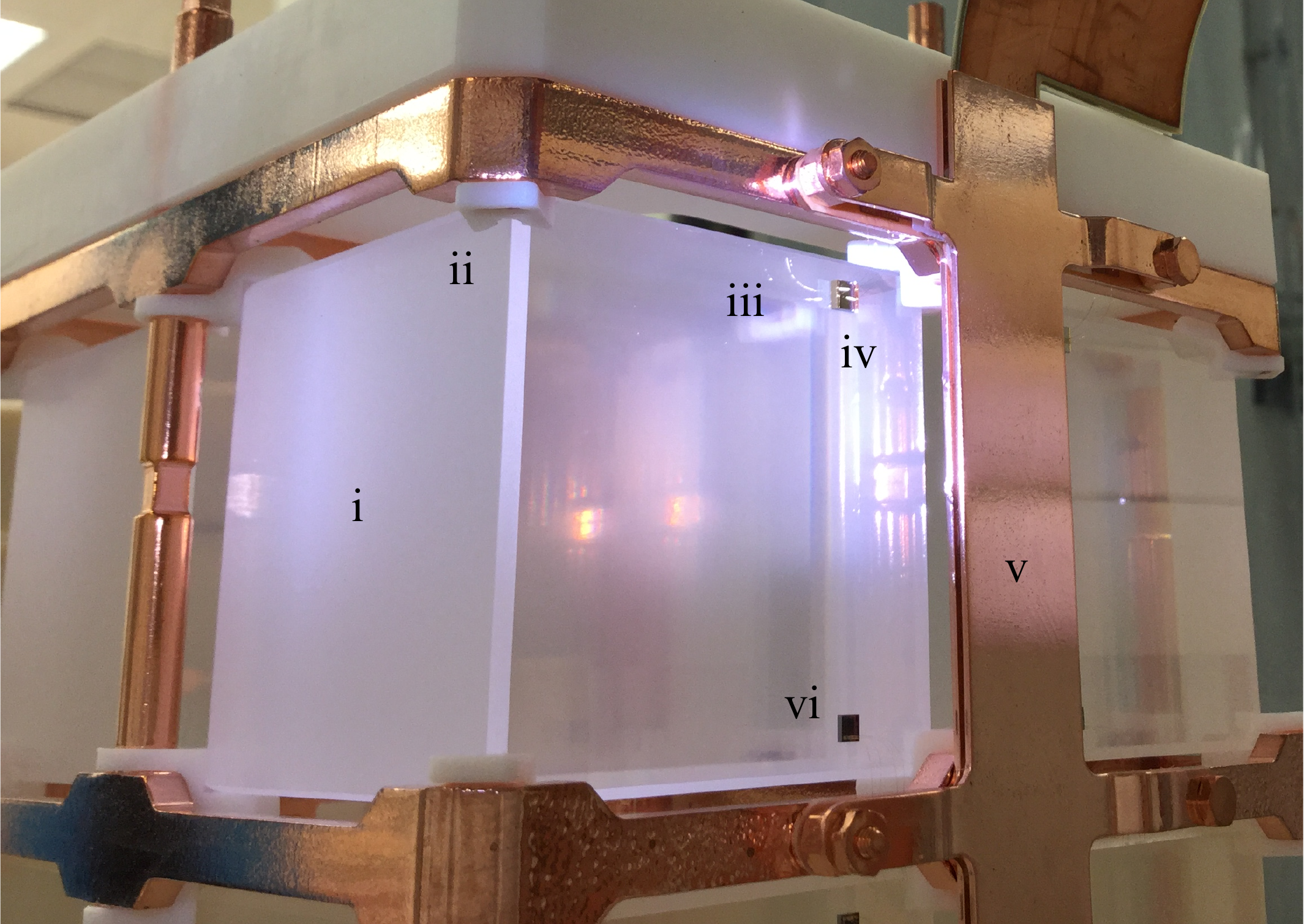}
    \caption{\textbf{CUORE calorimeter.} Shown are the absorber (i - TeO$_2$ crystal of $5\times5\times5$ cm$^3$ size); weak thermal couplings (ii - PTFE spacer, iii - Au wire); temperature sensor (iv - NTD Ge thermistor); heat sink (v - Cu frame); and auxiliary Joule heater (vi - Si chip). A spotlight was used to increase the visibility of the 25~$\mu$m 
    wires used for electrical connection.} 
    \label{fig:crystal}
\end{figure}

CUORE was designed to have each of the 988 calorimeters operate independently. In the CUORE calorimeter~(Fig.~\ref{fig:crystal}), the TeO$_2$ crystal acts as the absorber, and an affixed neutron transmutation doped (NTD) Ge thermistor~\cite{Haller1984} acts as the temperature sensor. Polytetrafluoroethylene (PTFE) spacers between the crystal and copper frame and gold wires, used for electrical connection of the NTD, provide weak thermal coupling~\cite{Alduino:2016vjd}. The copper frame structure acts as the heat sink given that it is in thermal contact with the cryostat mixing chamber, which sets the base temperature~\cite{Alduino:2019xia}. To correct our measurement against possible drifts in the operating temperature, a silicon heater is affixed to the crystal and used to induce periodic, monoenergetic pulses~\cite{Andreotti:2012zz,Carniti:2017zkr}. 
We measure energy depositions by applying an optimized constant bias current to the thermistor so that the voltage measured across 
it acts as a proxy for the temperature of the calorimeter~\cite{Arnaboldi:2017aek}. Owing to the combination of the thermal capacities and conductances of our calorimeter components, we measure slow pulses (i.e., $\sim$50~ms rise time, $\sim$1~s thermalization time) from the thermistor readout~\cite{256634}. Our continuous data stream is formed by passing the voltage though a room-temperature pre-amplifier and an anti-aliasing filter, and digitizing at 1~kHz~\cite{Arnaboldi:2010af,DiDomizio:2018ldc}.

\subsection*{Data collection, analysis pipeline, and event selection}

CUORE began collecting data in 2017 and made major strides toward stable detector operations during the early stages of data taking~\cite{CUORE:2021ctv}. Currently, the procedure to optimize all calorimeters is largely automated~\cite{Alfonso:2020yee} and routine cryogenic maintenance is minimal. Between 2019 and mid-2023, CUORE has been collecting data with $>$90\% duty cycle at a rate of $\sim$50~kg$\cdot$yr per month of TeO$_2$~\cite{CUOREPRLResult}. During this period, we were able to maintain steady exposure accumulation and operate the detector at different base temperatures. Thanks to the unprecedented stable operation of the cryogenic infrastructure, we collected over 2~tonne$\cdot$yr TeO$_2$ exposure.

CUORE data are organized into datasets. Each dataset begins and ends with $\sim$1~week of calibration measurements, during which $^{232}$Th or $^{232}$Th$+$$^{60}$Co radioactive sources are deployed around the detector. The high-intensity, characteristic gamma rays of well-known energies emitted by the radioactive sources are used to calibrate the voltage measurements in the entire dataset. Between the calibrations, we acquire $\sim$1.5~month-long physics data, which contribute to our reported TeO$_2$ exposure. The duration of the dataset maximizes physics data collection while ensuring that the calibration is consistent for all calorimeters and compatible with the detector conditions. 

\begin{figure}[H]
    \begin{subfigure}{1.0\linewidth}
    \includegraphics[width=1.0\textwidth]{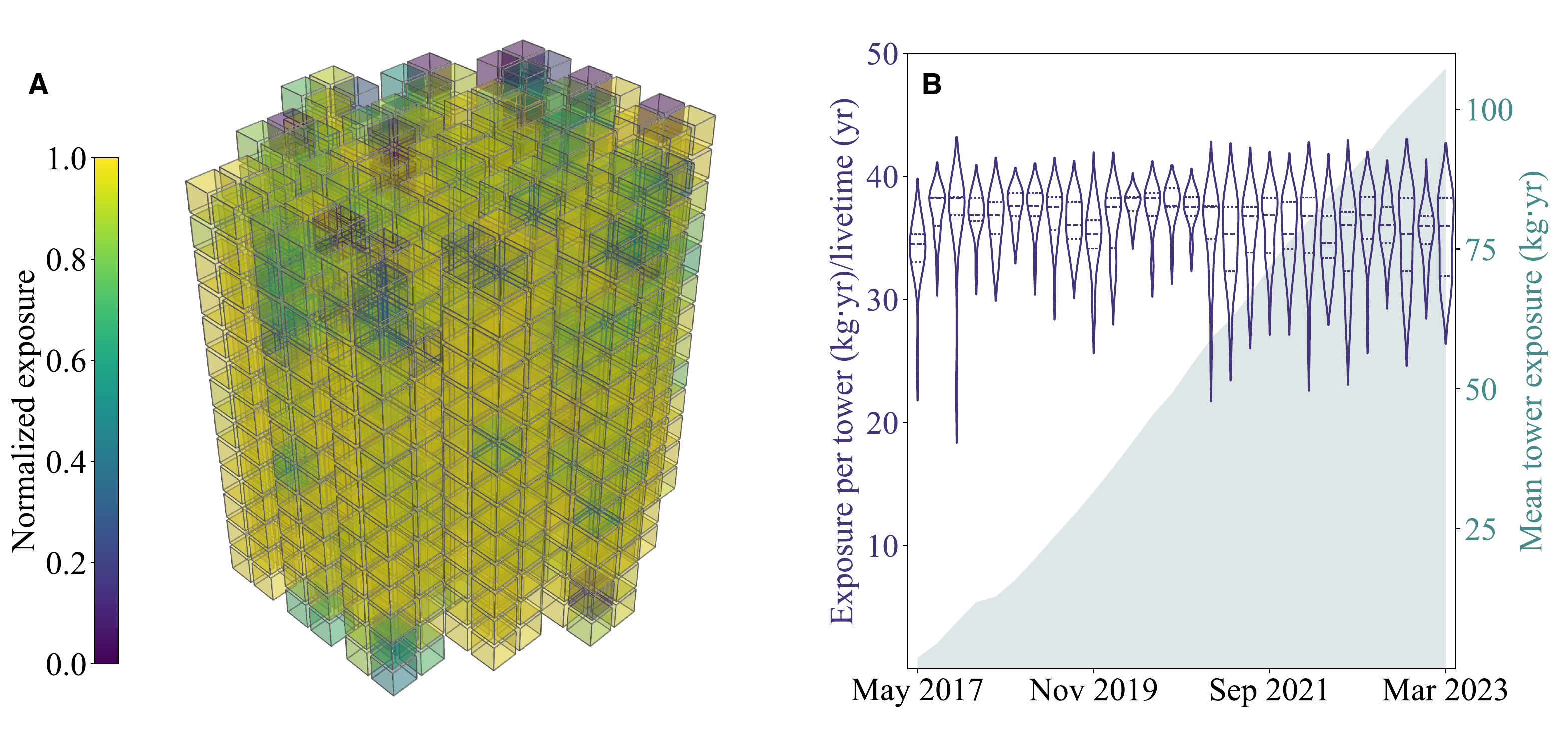}
        \caption{}
        \label{fig:CUORE_3d_exposure}
    \end{subfigure}\hfill
    \begin{subfigure}{1.0\linewidth}
        \caption{}
        \label{fig:exp_accumulation_time}
    \end{subfigure}
    \vspace*{-20mm}
    \caption{\textbf{CUORE exposure.} 
    (\textbf{A}) Schematic representation of the crystal array. The color of each crystal reflects its total analyzed exposure divided by the maximum value for all the calorimeters. The calorimeters near the top of their towers are more susceptible to vibrations. Because of their lower performance, they were removed from the analysis and consequently exhibit low exposures. (\textbf{B}) For each dataset, the distribution of the exposure collected on single towers normalized to the livetime of their active channels within the selected dataset is shown in purple. Each entry in the distribution is a proxy for the mass of data collected with a given tower based on the number of active channels available. The cumulative analyzed exposure averaged over towers by dataset is shown in green.}
\end{figure}

To search for the potential signal peak at \qbb, we identify pulses in the calorimeter readout that likely correspond to particle energy depositions and reconstruct their energies to build a spectrum. We processed 28 datasets and, after removing calorimeters performing below standard, our data processing framework reconstructed the energy spectra of an average of 914 out of 984 fully operational calorimeters per dataset. Because of the different operating conditions and intrinsic characteristics of the calorimeters, data acquisition is performed for each dataset-calorimeter independently. Hence, our energy reconstruction framework is largely automated and flexible to accommodate the number of calorimeters and their different characteristics~\cite{Alfonso:2020yee}. Figure~\ref{fig:CUORE_3d_exposure} demonstrates the uniformity of the analyzed exposure distribution over all 988 calorimeters, which highlights the effectiveness of our analysis framework and its ability to adapt to each calorimeter. The steady and uniform TeO$_2$ exposure accumulation of CUORE over time is shown in Fig.~\ref{fig:exp_accumulation_time}.

The typical response of CUORE calorimeters to particles impinging on the crystals is of order 100~$\mu$V/MeV of deposited energy. The frequency bandwidth of the corresponding pulses is 0 -- 20~Hz; the noise components extend to higher frequencies (fig.~\ref{fig:anps}). The main sources of environmental noise that can degrade our detector energy resolution are electrical interference, mechanical vibrations, and microphonic noise~\cite{Alduino:2019xia}. Microphones, low-frequency accelerometers, and a triaxial seismometer have been installed in the proximity of the cryostat to independently measure the noise, covering a frequency range between 0.1~Hz -- 1~kHz. For each dataset, a subset of these devices is used in our denoising algorithm~\cite{Vetter:2023fas}. New to this analysis, our denoising procedure combines the device measurements with the noise extracted from the calorimeter readout to build a multivariate nonlinear offline noise cancellation algorithm~\cite{supplementary}. 

To identify pulses and classify them as events for the energy reconstruction, we apply a low-threshold trigger, called optimal trigger (OT)~\cite{DiDomizio:2010ph}, to the denoised data. Then, to measure the pulse amplitude more accurately, we filter events with the 
optimal (matched) filter (OF)~\cite{Gatti:1986cw} that maximizes the signal-to-noise ratio by exploiting the distinct shapes of the stochastic noise and the particle-induced signal power spectra. Because fluctuations in the baseline temperature can affect the pulse amplitude thereby degrading our energy resolution, we use events corresponding to fixed energy depositions (i.e., heater-induced or gamma peak-based) to characterize and correct for this effect. To convert the corrected amplitudes to energy, we use calibration data to correlate the peak positions in the amplitude spectrum with the gamma-ray emissions of the deployed radioactive sources~\cite{Alduino:2017ehq}.

Our segmented detector design enables us to perform coincidence-based event selections. Monte Carlo simulations indicate that $\sim$88\% of \ndbd\ events are contained in a single crystal whereas events induced by crystal surface radioactivity and gamma rays undergoing Compton scattering could deposit energy in multiple crystals. Therefore, we remove events depositing $>$40~keV in more than one crystal within 5~ms of each other~\cite{CUORE:2020bok,CUORE:2020bok_erratum}. This time interval is a compromise between the time resolution of our detectors and the rate of random coincidences. 

As a way to select reliable \ndbd\ candidate events, we perform pulse shape discrimination~(PSD) to reject pulses with irregular shapes. For each event, we evaluate a Normalized Reconstruction Error~(NRE), a metric based on principal component analysis, to reject nonstandard pulses primarily caused by pileup, noise, and abrupt changes in operating conditions. An NRE threshold tuned to optimize the \ndbd\ sensitivity is used for our event selection~\cite{Huang:2020mko,CUORE:2021mvw}. The energy spectrum after applying the aforementioned event selections is shown in fig.~\ref{fig:spectrum}.

\subsection*{Detector response}

After selecting the events to build the spectrum for our search, we blind our data by exchanging a random fraction ($\sim$10$\%$) of events between the 2615~keV peak and \qbb\ to produce an artificial signal peak. Using the blinded data, we fit a model that reflects the background and the posited \qbb\ peak to an energy region of interest (ROI) of [2465,2575]~keV. The range is chosen to be wide enough to contain adequate background statistics while excluding unnecessary nuisance radioactive peaks. This procedure leads to an unbiased fit optimization. 

\begin{figure}[H]
    \centering
    \includegraphics[width=1.0\textwidth]{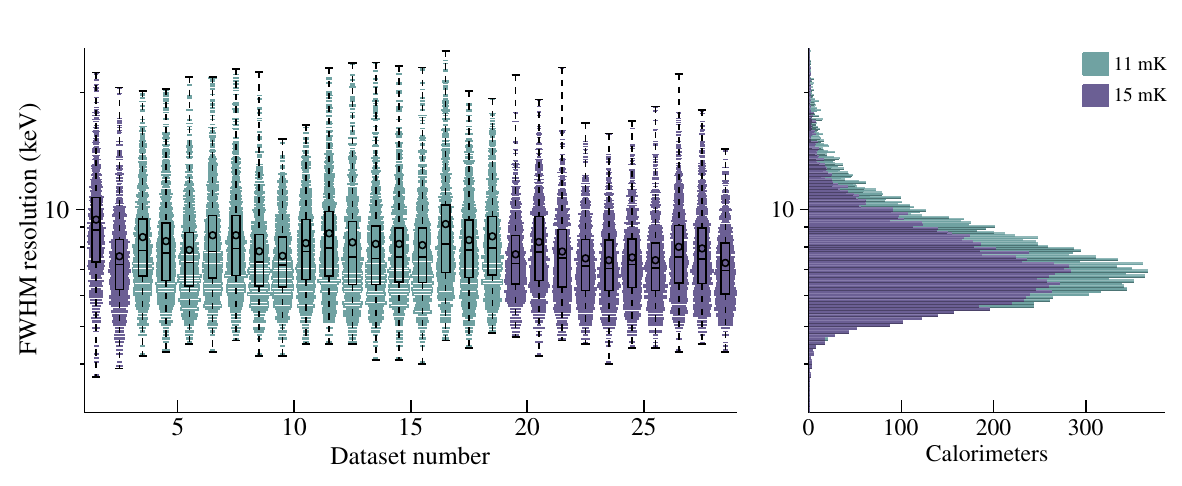}
    \caption{\textbf{Calorimeter energy resolution.} 
    Left: The distribution of the FWHM energy resolution of the CUORE active calorimeters at the $^{208}$Tl 2615~keV peak in calibration data for each dataset. The function drawn along the violin distribution represents the frequency of the FWHM energy resolution over the CUORE detector array. 
    The colors indicate the base temperature at the time of measurement: 11~mK in green and 15~mK in purple. Right: One-dimensional projections of the resolutions corresponding to each temperature.} 
    \label{fig:tl-resolution}
\end{figure}

To determine the detector response of the individual calorimeters in our model, we use gamma peaks in both calibration and physics data~\cite{supplementary}. We evaluate the expected shape of a monoenergetic peak at \qbb\ by fitting the high-statistics 2615~keV peak in calibration data (fig.~\ref{fig:ls_calibration}). Combining the fit results from all calorimeters and datasets, we obtain an exposure-weighted harmonic mean full width at half maximum (FWHM) energy resolution of (7.540~$\pm$~0.024)~keV at the 2615~keV calibration peak. Figure~\ref{fig:tl-resolution} shows the uniform spread of the FWHM resolution over all calorimeters and datasets, which underlines the efficacy of our analysis infrastructure and long-term reliability of our system. 

When fitting the peaks in physics data, we maintain the calorimeter-dependence of the detector response by parametrizing the peak resolution in terms of the calibration-based fit results~\cite{supplementary}. Using the extrapolation of the energy resolution to \qbb\ for each dataset-calorimeter, we report an exposure-weighted harmonic mean FWHM energy resolution of (7.310~$\pm$~0.024)~keV, corresponding to 0.3\% relative resolution. The calibration bias, defined as the difference between the reconstructed and nominal peak energy, is $\big($0.40$^{+0.21}_{-0.44}\big)$~keV at \qbb. 

\subsection*{Search for \ndbdbold}

After performing all selection cuts, the remaining data correspond to 2039.0 kg$\cdot$yr TeO$_2$~(567.0 kg$\cdot$yr $^{130}$Te) exposure. To search for \ndbd, we evaluate the number of events above background at \qbb---i.e., 2527.5~keV. 
We limit our search to the ROI which contains a total of 3504 events. The background in the ROI includes a peak at 2505.7~keV from the simultaneous absorption of two gamma rays from $^{60}$Co and a uniform background predominantly from degraded alpha particles from surface contamination, with a contribution from multi-Compton scattering of 2615~keV gamma rays~\cite{CUORE:2024fak}.

\begin{figure}[H]
    \centering
    \includegraphics[width=0.6\textwidth]{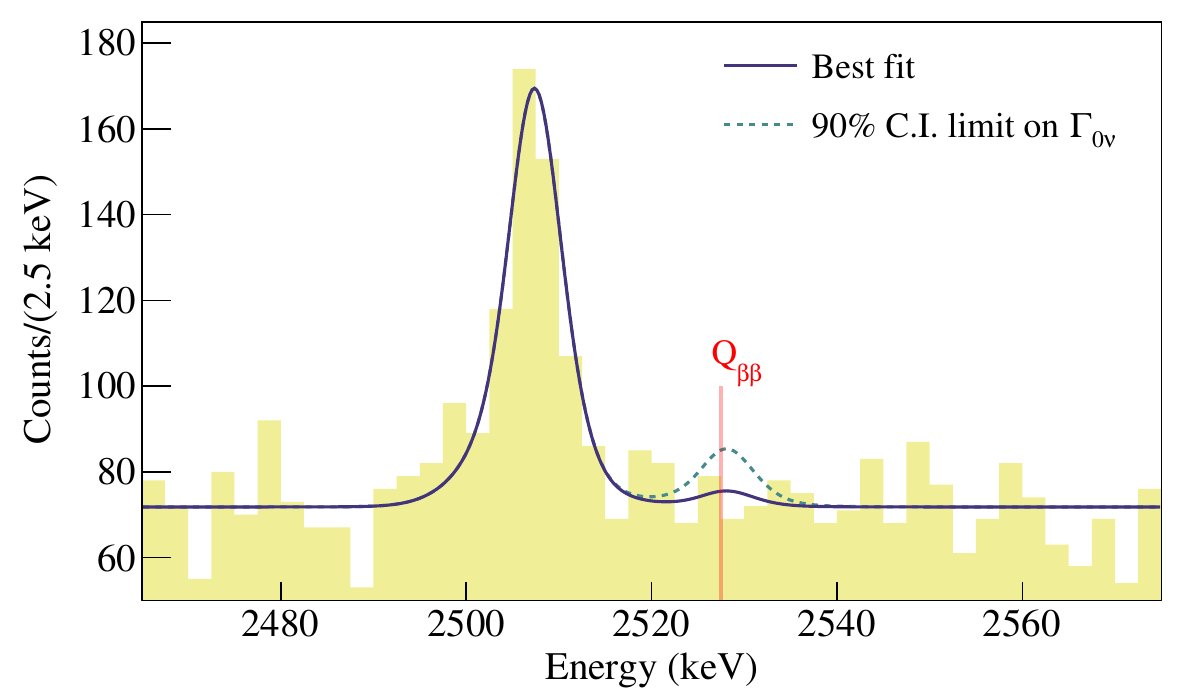} 
    \caption{
    \textbf{Fit to the ROI spectrum.} The best-fit curve (dark purple) and the best-fit curve with the \ndbd\ rate fixed to the 90\%~C.I. limit (dashed green) to the spectrum (yellow) in the ROI after all selection cuts.}
    \label{fig:fit}
\end{figure}

We model the spectrum in the ROI as the sum of a \qbb\ peak, a dataset-dependent uniform background, and a time-dependent $^{60}$Co sum peak and perform an unbinned Bayesian fit. We use our dataset-calorimeter evaluation of the detector efficiencies and detector response as priors in the fit, where the detection efficiency is evaluated as the product of the containment efficiency, the reconstruction efficiency, the anti-coincidence efficiency, and the pulse shape discrimination efficiency~\cite{supplementary}. Their values are summarized in table~\ref{tab:Efficiencies}. Systematic effects are included as additional nuisance parameters~\cite{supplementary}. The fit to the ROI is shown in Fig.~\ref{fig:fit}.

We observe no statistically significant evidence of \ndbd\ and report a best-fit decay rate of $\hat{\Gamma}_{0 \nu} = 5.5^{+7.3}_{-5.5}$ (stat.+syst.) $\times$ 10$^{-27}$~yr$^{-1}$ and set a limit on the decay rate of $\Gamma^{0\nu}_{1/2} <$ 2.0 $\times$ 10$^{-26}$~yr$^{-1}$ at 90\% credibility interval (C.I.) corresponding to a decay half-life limit of $T^{0\nu}_{1/2}$ $>$ 3.5 $\times$ 10$^{25}$~yr (90\% C.I.)~(fig.~\ref{fig:posterior}). We also perform a frequentist fit and, using the profile likelihood ratio test statistic~\cite{Algeri:2019lah}, set a half-life limit of $T^{0\nu}_{1/2}$ $>$ 3.4 $\times$ 10$^{25}$~yr at 90\% confidence level (C.L.)~(fig.~\ref{fig:posteriorNLL}), which is compatible with the Bayesian fit result. Using the background-only hypothesis to fit the ROI, we obtain a background index~(BI) for each dataset (fig.~\ref{fig:BI}) with an exposure-weighted mean of  $\big($1.42$^{+0.03}_{-0.02}$$\big)$ $\times$ 10$^{-2}$~counts~keV$^{-1}$kg$^{-1}$yr$^{-1}$. We use the BIs to extract a median exclusion sensitivity of 4.4 $\times$ 10$^{25}$~yr (90\% C.I.) (fig.~\ref{fig:excl_sens})~\cite{supplementary}. Compared to this value, the probability of obtaining a stronger limit is 74\%. 

\begin{figure}[H]
    \centering
    \includegraphics[width=0.6\textwidth]{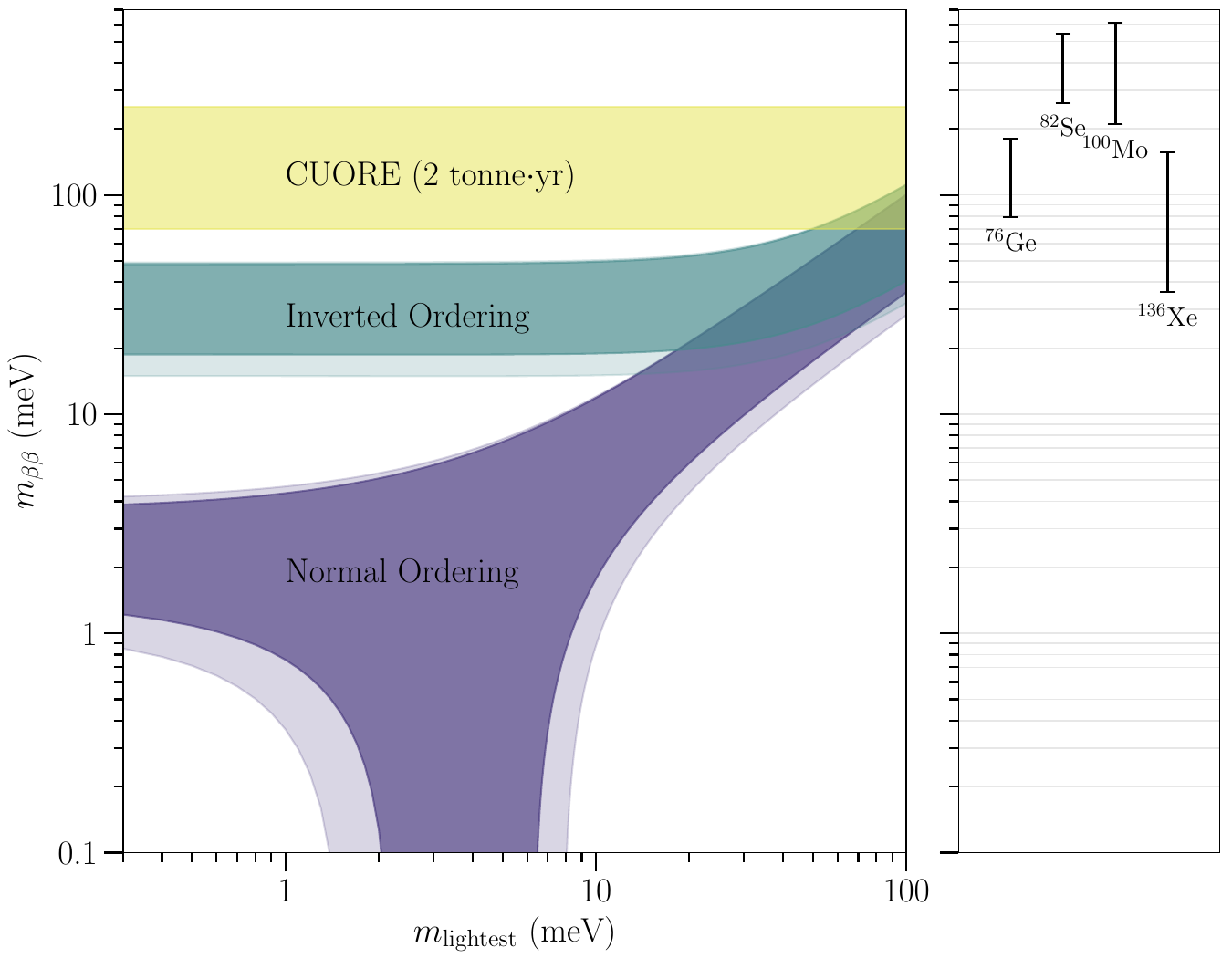} 
    \caption{
    \textbf{Limits on \mbbbold.} The allowed parameter space as a function of the lightest neutrino mass in the case of inverted (normal) ordering is shown in green (purple). The lighter shaded areas correspond to the 3$\sigma$ uncertainties on the oscillation parameters~\cite{Agostini:2022zub}. The yellow band corresponds to the limit obtained from this analysis. Limits obtained from $^{76}$Ge~\cite{GERDA:PRL}, 
    $^{82}$Se~\cite{CUPID:2022puj}, 
    $^{100}$Mo~\cite{Agrawal:2025}, 
    $^{136}$Xe~\cite{KamLAND-Zen:2022tow} are also shown.} 
    \label{fig:lobster}
\end{figure}

The simplest model that is commonly used to compare \ndbd\ search results across isotopes is the light-neutrino exchange model~\cite{Agostini:2022zub} from which we extract an effective Majorana neutrino mass (\mbb). Thus, we place a limit of \mbb\ $<$ 70 -- 250~meV, where the spread comes from the range of different nuclear model parameters currently available in the literature~\cite{Vaquero:2014dna,Hyvarinen:2015bda,Horoi:2016sm,Song:2017prc,Menendez:2018jpg,Fang:2018prc,Simkovic:2018prc,Coraggio:2020prc,Deppisch:2020prd}. The CUORE limit, along with the most stringent \mbb\ constraints obtained with other isotopes are shown in Fig.~\ref{fig:lobster}. 

\subsection*{Discussion and outlook}

CUORE has demonstrated over 5 years stable operation of an array of 988 cryogenic calorimeters and will continue taking data until reaching an analyzed exposure of 3~tonne$\cdot$yr TeO$_2$ ($\sim$1~tonne$\cdot$yr $^{130}$Te). The instrumentation and the analysis infrastructure developed in this work not only enables a world-leading \ndbd\ search, but also establishes large-volume cryogenic calorimetry as a unique path for a future multi-isotope \ndbd\ search program. After CUORE reaches its target exposure, the experiment will begin phase II and pivot its primary focus to low-energy processes.

CUPID (CUORE Upgrade with Particle IDentification)~\cite{CUPIDInterestGroup:2019inu}, a next-generation experiment that will search for \ndbd\ in $^{100}$Mo, will follow CUORE phase II. The CUPID detector is designed to improve upon our \mbb\ sensitivity by a factor of 5, deeply probing the inverted mass ordering parameter space assuming the light-neutrino exchange model. If CUPID discovers \ndbd, it would confirm the Majorana nature of the neutrino and compel precise measurements of \ndbd\ across multiple isotopes making cryogenic calorimeters all the more relevant. 
Because CUPID will use the same detection technique and will be hosted in the CUORE cryogenic facility, the experiment will inherit a well-understood system and have working knowledge of the cryostat operations, background, and noise from CUORE. CUORE has thus laid the groundwork not only for CUPID, but other future large-array cryogenic-calorimeter-based experiments.

\newpage


\clearpage 

%
\bibliography{science_template} 

\begin{thebibliography}{10}
\providecommand{\url}[1]{\texttt{#1}}
\expandafter\ifx\csname urlstyle\endcsname\relax
  \providecommand{\doi}[1]{doi:\discretionary{}{}{}#1}\else
  \providecommand{\doi}{doi:\discretionary{}{}{}\begingroup
  \urlstyle{rm}\Url}\fi

\bibitem{ParticleDataGroup:2022pth}
R.~L. Workman, \emph{et~al.}, {Review of Particle Physics}. \emph{PTEP}
  \textbf{2022}, 083C01 (2022), \url{https://doi.org/10.1093/ptep/ptac097}.

\bibitem{Dine:2003ax}
M.~Dine, A.~Kusenko, {The Origin of the matter - antimatter asymmetry}.
  \emph{Rev. Mod. Phys.} \textbf{76}, 1 (2003),
  \url{https://doi.org/10.1103/RevModPhys.76.1}.

\bibitem{Majorana:1937vz}
E.~Majorana, {Theory of the Symmetry of Electrons and Positrons (in Italian)}.
  \emph{Nuovo Cim.} \textbf{14}, 171--184 (1937),
  \url{https://doi.org/10.1007/BF02961314}.

\bibitem{GoeppertMayer:1935qp}
M.~Goeppert-Mayer, {Double beta-disintegration}. \emph{Phys.\ Rev.}
  \textbf{48}, 512--516 (1935), \url{https://doi.org/10.1103/PhysRev.48.512}.

\bibitem{Furry:1939qr}
W.~H. Furry, {On transition probabilities in double beta-disintegration}.
  \emph{Phys.\ Rev.} \textbf{56}, 1184--1193 (1939),
  \url{https://doi.org/10.1103/PhysRev.56.1184}.

\bibitem{Sakharov:1967dj}
A.~D. Sakharov, {Violation of CP Invariance, C asymmetry, and baryon asymmetry
  of the universe}. \emph{Pisma Zh. Eksp. Teor. Fiz.} \textbf{5}, 32--35
  (1967), \url{https://dx.doi.org/10.1070/PU1991v034n05ABEH002497}.

\bibitem{Augier:2023prl}
C.~Augier, \emph{et~al.}, Measurement of the
  $2\ensuremath{\nu}\ensuremath{\beta}\ensuremath{\beta}$ Decay Rate and
  Spectral Shape of $^{100}\mathrm{Mo}$ from the CUPID-Mo Experiment.
  \emph{Phys. Rev. Lett.} \textbf{131}, 162501 (2023),
  \url{https://link.aps.org/doi/10.1103/PhysRevLett.131.162501}.

\bibitem{CUPID:2023wyy}
O.~Azzolini, \emph{et~al.}, {Measurement of the
  2\ensuremath{\nu}\ensuremath{\beta}\ensuremath{\beta} Decay Half-Life of
  $^{82}\mathrm{Se}$ with the Global CUPID-0 Background Model}. \emph{Phys.
  Rev. Lett.} \textbf{131}~(22), 222501 (2023),
  \url{https://doi.org/10.1103/PhysRevLett.131.222501}.

\bibitem{CUORE:2020bok}
D.~Q. Adams, \emph{et~al.}, Measurement of the $2\nu\beta\beta$ Decay Half-Life
  of $^{130}\mathrm{Te}$ with {CUORE}. \emph{Phys. Rev. Lett.}
  \textbf{126}~(17), 171801 (2021),
  \url{https://doi.org/10.1103/PhysRevLett.126.171801}.

\bibitem{CUORE:2020bok_erratum}
D.~Q. Adams, \emph{et~al.}, Erratum: Measurement of the $2\nu\beta\beta$ Decay
  Half-Life of $^{130}\mathrm{Te}$ with {CUORE}. \emph{Phys. Rev. Lett.}
  \textbf{131}~(24), 249902 (2023),
  \url{https://doi.org/10.1103/PhysRevLett.131.249902}.

\bibitem{Dolinski:Review}
M.~J. Dolinski, A.~W. Poon, W.~Rodejohann, Neutrinoless Double-Beta Decay:
  Status and Prospects. \emph{Ann. Rev. Nucl. Part. Sci.} \textbf{69}~(1),
  219--251 (2019), \url{https://doi.org/10.1146/annurev-nucl-101918-023407}.

\bibitem{Agostini:2022zub}
M.~Agostini, G.~Benato, J.~A. Detwiler, J.~Men\'endez, F.~Vissani, {Toward the
  discovery of matter creation with neutrinoless
  \ensuremath{\beta}\ensuremath{\beta} decay}. \emph{Rev. Mod. Phys.}
  \textbf{95}~(2), 025002 (2023),
  \url{https://doi.org/10.1103/RevModPhys.95.025002}.

\bibitem{Fiorini:1983yj}
E.~Fiorini, T.~O. Niinikoski, Low Temperature Calorimetry for Rare Decays.
  \emph{Nucl. Instrum. Meth. A} \textbf{224}, 83 (1984),
  \url{https://doi.org/10.1016/0167-5087(84)90449-6}.

\bibitem{Brofferio:2018lys}
C.~Brofferio, S.~Dell'Oro, Contributed Review: The saga of neutrinoless double
  beta decay search with {TeO}$_2$ thermal detectors. \emph{Rev. Sci. Instrum.}
  \textbf{89}~(12), 121502 (2018), \url{https://doi.org/10.1063/1.5031485}.

\bibitem{CUORE:2021ctv}
D.~Q. Adams, \emph{et~al.}, {CUORE opens the door to tonne-scale cryogenics
  experiments}. \emph{Prog. Part. Nucl. Phys.} \textbf{122}, 103902 (2022),
  \url{https://doi.org/10.1016/j.ppnp.2021.103902}.

\bibitem{CUORE:2021mvw}
D.~Q. Adams, \emph{et~al.}, {Search for Majorana neutrinos exploiting
  millikelvin cryogenics with CUORE}. \emph{Nature} \textbf{604}~(7904), 53--58
  (2022), \url{https://doi.org/10.1038/s41586-022-04497-4}.

\bibitem{CUPID:2022puj}
O.~Azzolini, \emph{et~al.}, {Final Result on the Neutrinoless Double Beta Decay
  of $^{82}$Se with CUPID-0}. \emph{Phys. Rev. Lett.} \textbf{129}~(11), 111801
  (2022), \url{https://doi.org/10.1103/PhysRevLett.129.111801}.

\bibitem{Augier:2022fr}
C.~Augier, \emph{et~al.}, Final results on the $0\nu\beta\beta$ decay half-life
  limit of $^{100}$Mo from the CUPID-Mo experiment. \emph{Eur. Phys. J. C}
  \textbf{82}~(11), 1033 (2022),
  \url{https://doi.org/10.1140/epjc/s10052-022-10942-5}.

\bibitem{Agrawal:2025}
A.~Agrawal, \emph{et~al.}, Improved Limit on Neutrinoless Double Beta Decay of
  $^{100}\mathrm{Mo}$ from AMoRE-I. \emph{Phys. Rev. Lett.} \textbf{134},
  082501 (2025), \doi{10.1103/PhysRevLett.134.082501},
  \url{https://link.aps.org/doi/10.1103/PhysRevLett.134.082501}.

\bibitem{Cremonesi:2013vla}
O.~Cremonesi, M.~Pavan, {Challenges in Double Beta Decay}. \emph{Adv.\ High
  Energy Phys.} \textbf{2014}, 951432 (2013),
  \url{https://doi.org/10.1155/2014/951432}.

\bibitem{Armengaud:2017cry}
E.~Armengaud, \emph{et~al.}, Development of $^{100}$Mo-containing scintillating
  bolometers for a high-sensitivity neutrinoless double-beta decay search.
  \emph{Eur. Phys. J. C} \textbf{77}~(785), 785 (2017),
  \url{https://doi.org/10.1140/epjc/s10052-017-5343-2}.

\bibitem{ARMATOL2024169936}
A.~Armatol, \emph{et~al.}, BINGO innovative assembly for background reduction
  in bolometric \ndbd\ experiments. \emph{Nucl. Instrum. Meth. A}
  \textbf{1069}, 169936 (2024),
  \url{https://doi.org/10.1016/j.nima.2024.169936}.

\bibitem{PhysRevApplied.20.064017}
V.~Singh, \emph{et~al.}, Large-area photon calorimeter with Ir-Pt bilayer
  transition-edge sensor for the CUPID experiment. \emph{Phys. Rev. Applied}
  \textbf{20}, 064017 (2023),
  \url{https://link.aps.org/doi/10.1103/PhysRevApplied.20.064017}.

\bibitem{Giuliani:2017tqo}
A.~Giuliani, F.~A. Danevich, V.~I. Tretyak, {A multi-isotope $0\nu2\beta$
  bolometric experiment}. \emph{Eur. Phys. J. C} \textbf{78}~(3), 272 (2018),
  \url{https://doi.org/10.1140/epjc/s10052-018-5750-z}.

\bibitem{Alduino:2019xia}
C.~Alduino, \emph{et~al.}, {The CUORE cryostat: An infrastructure for rare
  event searches at millikelvin temperatures}. \emph{Cryogenics} \textbf{102},
  9--21 (2019), \url{https://doi.org/10.1016/j.cryogenics.2019.06.011}.

\bibitem{CUORE:2021xns}
D.~Q. Adams, \emph{et~al.}, {Search for double-beta decay of $\mathrm
  {^{130}Te}$ to the $0^+$ states of $\mathrm {^{130}Xe}$ with CUORE}.
  \emph{Eur. Phys. J. C} \textbf{81}~(7), 567 (2021),
  \url{https://doi.org/10.1140/epjc/s10052-021-09317-z}.

\bibitem{CUORE120Te}
D.~Q. Adams, \emph{et~al.}, Search for neutrinoless
  ${\ensuremath{\beta}}^{+}$EC decay of $^{120}\mathrm{Te}$ with CUORE.
  \emph{Phys. Rev. C} \textbf{105}, 065504 (2022),
  \url{https://link.aps.org/doi/10.1103/PhysRevC.105.065504}.

\bibitem{CUORE128Te}
D.~Q. Adams, \emph{et~al.}, New Direct Limit on Neutrinoless Double Beta Decay
  Half-Life of $^{128}\mathrm{Te}$ with CUORE. \emph{Phys. Rev. Lett.}
  \textbf{129}, 222501 (2022),
  \url{https://link.aps.org/doi/10.1103/PhysRevLett.129.222501}.

\bibitem{Rahaman:2011zz}
S.~Rahaman, \emph{et~al.}, {Double-beta decay Q values of $^{116}$Cd and
  $^{130}$Te}. \emph{Phys. Lett. B} \textbf{703}, 412--416 (2011),
  \url{https://doi.org/10.1016/j.physletb.2011.07.078}.

\bibitem{Borexino:2012wej}
G.~Bellini, \emph{et~al.}, {Cosmic-muon flux and annual modulation in Borexino
  at 3800 m water-equivalent depth}. \emph{J.\ Cosmol.\ Astropart.\ Phys.}
  \textbf{05}, 015 (2012), \url{https://doi.org/10.1088/1475-7516/2012/05/015}.

\bibitem{Wulandari:2003cr}
H.~Wulandari, J.~Jochum, W.~Rau, F.~von Feilitzsch, {Neutron flux at the Gran
  Sasso underground laboratory revisited}. \emph{Astropart.\ Phys.}
  \textbf{22}, 313--322 (2004),
  \url{https://doi.org/10.1016/j.astropartphys.2004.07.005}.

\bibitem{Pattavina:2019pxw}
L.~Pattavina, \emph{et~al.}, Radiopurity of an archeological Roman Lead
  cryogenic detector. \emph{Eur. Phys. J. A} \textbf{55}, 127 (2019),
  \url{https://doi.org/10.1140/epja/i2019-12809-0}.

\bibitem{CUORE:2024fak}
D.~Q. Adams, \emph{et~al.}, {Data-driven background model for the CUORE
  experiment}. \emph{Phys. Rev. D} \textbf{110}~(5), 052003 (2024),
  \url{https://doi.org/10.1103/PhysRevD.110.052003}.

\bibitem{Alessandria:2012zp}
F.~Alessandria, \emph{et~al.}, {Validation of techniques to mitigate copper
  surface contamination in CUORE}. \emph{Astropart.\ Phys.} \textbf{45}, 13--22
  (2013), \url{https://doi.org/10.1016/j.astropartphys.2013.02.005}.

\bibitem{Buccheri:2014bma}
E.~Buccheri, \emph{et~al.}, {An assembly line for the construction of
  ultra-radio-pure detectors}. \emph{Nucl.\ Instrum.\ Meth.\ A} \textbf{768},
  130--140 (2014), \url{https://doi.org/10.1016/j.nima.2014.09.046}.

\bibitem{Benato:2017kdf}
G.~Benato, \emph{et~al.}, Radon mitigation during the installation of the
  {CUORE} 0$\nu\beta\beta$ decay detector. \emph{J. Instrum.} \textbf{13}~(01),
  P01010 (2018), \url{https://dx.doi.org/10.1088/1748-0221/13/01/P01010}.

\bibitem{Haller1984}
E.~E. Haller, N.~P. Palaio, M.~Rodder, W.~L. Hansen, E.~Kreysa, \emph{NTD
  Germanium: A Novel Material for Low Temperature Bolometers} (Springer US,
  Boston, MA), pp. 21--36 (1984),
  \url{https://doi.org/10.1007/978-1-4613-2695-3_2}.

\bibitem{Alduino:2016vjd}
C.~Alduino, \emph{et~al.}, {CUORE}-0 detector: design, construction and
  operation. \emph{J. Instrum.} \textbf{11}~(07), P07009 (2016),
  \url{https://dx.doi.org/10.1088/1748-0221/11/07/P07009}.

\bibitem{Andreotti:2012zz}
E.~Andreotti, \emph{et~al.}, Production, characterization, and selection of the
  heating elements for the response stabilization of the {CUORE} bolometers.
  \emph{Nucl. Instrum. Meth. A} \textbf{664}, 161--170 (2012),
  \url{https://doi.org/10.1016/j.nima.2011.10.065}.

\bibitem{Carniti:2017zkr}
K.~Alfonso, \emph{et~al.}, A High Precision Pulse Generation and Stabilization
  System for Bolometric Experiments. \emph{J. Instrum.} \textbf{13}~(02),
  P02029 (2018), \url{https://dx.doi.org/10.1088/1748-0221/13/02/P02029}.

\bibitem{Arnaboldi:2017aek}
C.~Arnaboldi, \emph{et~al.}, A front-end electronic system for large arrays of
  bolometers. \emph{J. Instrum.} \textbf{13}~(02), P02026 (2018),
  \url{https://dx.doi.org/10.1088/1748-0221/13/02/P02026}.

\bibitem{256634}
A.~Alessandrello, \emph{et~al.}, An electrothermal model for large mass
  bolometric detectors. \emph{IEEE Transactions on Nuclear Science}
  \textbf{40}~(4), 649--656 (1993), \url{https://dx.doi.org/10.1109/23.256634}.

\bibitem{Arnaboldi:2010af}
C.~Arnaboldi, M.~Cariello, S.~{Di Domizio}, A.~Giachero, G.~Pessina, A
  programmable multichannel antialiasing filter for the CUORE experiment.
  \emph{Nucl. Instrum. Meth. A} \textbf{617}~(1), 327--328 (2010),
  \url{https://doi.org/10.1016/j.nima.2009.09.023}.

\bibitem{DiDomizio:2018ldc}
S.~Di~Domizio, \emph{et~al.}, A data acquisition and control system for large
  mass bolometer arrays. \emph{J. Instrum.} \textbf{13}~(12), P12003 (2018),
  \url{https://dx.doi.org/10.1088/1748-0221/13/12/P12003}.

\bibitem{Alfonso:2020yee}
K.~Alfonso, \emph{et~al.}, {An automated system to define the optimal operating
  settings of cryogenic calorimeters}. \emph{Nucl. Instrum. Meth. A}
  \textbf{1008}, 165451 (2021),
  \url{https://doi.org/10.1016/j.nima.2021.165451}.

\bibitem{CUOREPRLResult}
D.~Q. Adams, \emph{et~al.}, Improved Limit on Neutrinoless Double-Beta Decay in
  $^{130}${Te} with {CUORE}. \emph{Phys. Rev. Lett.} \textbf{124}~(12), 122501
  (2020), \url{https://doi.org/10.1103/PhysRevLett.124.122501}.

\bibitem{Vetter:2023fas}
K.~J. Vetter, \emph{et~al.}, {Improving the performance of cryogenic
  calorimeters with nonlinear multivariate noise cancellation algorithms}.
  \emph{Eur. Phys. J. C} \textbf{84}~(3), 243 (2024),
  \url{https://doi.org/10.1140/epjc/s10052-024-12595-y}.

\bibitem{supplementary}
{Materials and methods are available as supplementary materials.}

\bibitem{DiDomizio:2010ph}
S.~Di~Domizio, F.~Orio, M.~Vignati, Lowering the energy threshold of large-mass
  bolometric detectors. \emph{J. Instrum.} \textbf{6}, P02007 (2011),
  \url{https://dx.doi.org/10.1088/1748-0221/6/02/P02007}.

\bibitem{Gatti:1986cw}
E.~Gatti, P.~F. Manfredi, Processing the Signals From Solid State Detectors in
  Elementary Particle Physics. \emph{Riv. Nuovo Cim.} \textbf{9}, 1--146
  (1986), \url{https://doi.org/10.1007/BF02822156}.

\bibitem{Alduino:2017ehq}
C.~Alduino, \emph{et~al.}, {First Results from CUORE: A Search for Lepton
  Number Violation via $0\nu\beta\beta$ Decay of $^{130}$Te}. \emph{Phys.\
  Rev.\ Lett.} \textbf{120}~(13), 132501 (2018),
  \url{https://doi.org/10.1103/PhysRevLett.120.132501}.

\bibitem{Huang:2020mko}
R.~Huang, \emph{et~al.}, Pulse shape discrimination in {CUPID-Mo} using
  principal component analysis. \emph{J. Instrum.} \textbf{16}~(03), P03032
  (2021), \url{https://doi.org/10.1088/1748-0221/16/03/p03032}.

\bibitem{Algeri:2019lah}
S.~Algeri, J.~Aalbers, K.~Dundas~Mor\r{a}, J.~Conrad, {Searching for new
  phenomena with profile likelihood ratio tests}. \emph{Nature Rev. Phys.}
  \textbf{2}~(5), 245--252 (2020),
  \url{https://doi.org/10.1038/s42254-020-0169-5}.

\bibitem{Vaquero:2014dna}
N.~L\'opez~Vaquero, T.~R. Rodr\'iguez, J.~L. Egido, Shape and pairing
  fluctuations effects on neutrinoless double beta decay nuclear matrix
  elements. \emph{Phys. Rev. Lett.} \textbf{111}~(14), 142501 (2013),
  \url{https://doi.org/10.1103/PhysRevLett.111.142501}.

\bibitem{Hyvarinen:2015bda}
J.~Hyv{\"a}rinen, J.~Suhonen, {Nuclear matrix elements for $0\nu\beta\beta$
  decays with light or heavy Majorana-neutrino exchange}. \emph{Phys. Rev.}
  \textbf{C91}~(2), 024613 (2015),
  \url{https://doi.org/10.1103/PhysRevC.91.024613}.

\bibitem{Horoi:2016sm}
M.~Horoi, A.~Neacsu, Shell model predictions for $^{124}\mathrm{Sn}$
  double-$\ensuremath{\beta}$ decay. \emph{Phys. Rev. C} \textbf{93}, 024308
  (2016), \url{https://link.aps.org/doi/10.1103/PhysRevC.93.024308}.

\bibitem{Song:2017prc}
L.~S. Song, J.~M. Yao, P.~Ring, J.~Meng, Nuclear matrix element of neutrinoless
  double-$\ensuremath{\beta}$ decay: Relativity and short-range correlations.
  \emph{Phys. Rev. C} \textbf{95}, 024305 (2017),
  \url{https://link.aps.org/doi/10.1103/PhysRevC.95.024305}.

\bibitem{Menendez:2018jpg}
J.~Menéndez, Neutrinoless $\beta \beta$ decay mediated by the exchange of
  light and heavy neutrinos: the role of nuclear structure correlations.
  \emph{Journal of Physics G: Nuclear and Particle Physics} \textbf{45}~(1),
  014003 (2017), \url{https://dx.doi.org/10.1088/1361-6471/aa9bd4}.

\bibitem{Fang:2018prc}
D.-L. Fang, A.~Faessler, F.~\ifmmode~\check{S}\else \v{S}\fi{}imkovic,
  $0\ensuremath{\nu}\ensuremath{\beta}\ensuremath{\beta}$-decay nuclear matrix
  element for light and heavy neutrino mass mechanisms from deformed
  quasiparticle random-phase approximation calculations for $^{76}\mathrm{Ge},
  ^{82}\mathrm{Se}, ^{130}\mathrm{Te}, ^{136}\mathrm{Xe}$, and
  $^{150}\mathrm{Nd}$ with isospin restoration. \emph{Phys. Rev. C}
  \textbf{97}, 045503 (2018),
  \url{https://link.aps.org/doi/10.1103/PhysRevC.97.045503}.

\bibitem{Simkovic:2018prc}
F.~\ifmmode~\check{S}\else \v{S}\fi{}imkovic, A.~Smetana, P.~Vogel,
  $0\ensuremath{\nu}\ensuremath{\beta}\ensuremath{\beta}$ and
  $2\ensuremath{\nu}\ensuremath{\beta}\ensuremath{\beta}$ nuclear matrix
  elements evaluated in closure approximation, neutrino potentials and SU(4)
  symmetry. \emph{Phys. Rev. C} \textbf{98}, 064325 (2018),
  \url{https://link.aps.org/doi/10.1103/PhysRevC.98.064325}.

\bibitem{Coraggio:2020prc}
L.~Coraggio, A.~Gargano, N.~Itaco, R.~Mancino, F.~Nowacki, Calculation of the
  neutrinoless double-$\ensuremath{\beta}$ decay matrix element within the
  realistic shell model. \emph{Phys. Rev. C} \textbf{101}, 044315 (2020),
  \url{https://link.aps.org/doi/10.1103/PhysRevC.101.044315}.

\bibitem{Deppisch:2020prd}
F.~F. Deppisch, L.~Graf, F.~Iachello, J.~Kotila, Analysis of light neutrino
  exchange and short-range mechanisms in
  $0\ensuremath{\nu}\ensuremath{\beta}\ensuremath{\beta}$ decay. \emph{Phys.
  Rev. D} \textbf{102}, 095016 (2020),
  \url{https://link.aps.org/doi/10.1103/PhysRevD.102.095016}.

\bibitem{CUPIDInterestGroup:2019inu}
W.~R. Armstrong, \emph{et~al.}, {CUPID pre-CDR} (2019),
  \url{https://doi.org/10.48550/arXiv.1907.09376}.

\bibitem{GERDA:PRL}
M.~Agostini, \emph{et~al.}, Final Results of {GERDA} on the Search for
  Neutrinoless Double-$\beta$ Decay. \emph{Phys. Rev. Lett.} \textbf{125},
  252502 (2020), \url{https://link.aps.org/doi/10.1103/PhysRevLett.125.252502}.

\bibitem{KamLAND-Zen:2022tow}
S.~Abe, \emph{et~al.}, {Search for the Majorana Nature of Neutrinos in the
  Inverted Mass Ordering Region with KamLAND-Zen}. \emph{Phys. Rev. Lett.}
  \textbf{130}~(5), 051801 (2023),
  \url{https://doi.org/10.1103/PhysRevLett.130.051801}.

\bibitem{zenodo}
P.~T. Surukuchi, C.~Bucci, K.~Alfonso, A.~Campani, Dataset for Constraints on
  Lepton Number Violation with the 2 tonne·yr CUORE Dataset. \emph{Zenodo}
  (2025), \doi{10.5281/zenodo.15742646},
  \url{https://doi.org/10.5281/zenodo.15742646}.

\bibitem{BATSoftware}
A.~Caldwell, D.~Koll\'{a}r, K.~Kr\"{o}ninger, {BAT - The Bayesian analysis
  toolkit}. \emph{Computer Physics Communications} \textbf{180}, 2197--2209
  (2009), \url{https://doi.org/10.1016/j.cpc.2009.06.026}.

\bibitem{GEANT4:2002zbu}
S.~Agostinelli, \emph{et~al.}, {GEANT4--a simulation toolkit}. \emph{Nucl.
  Instrum. Meth. A} \textbf{506}, 250--303 (2003),
  \url{https://doi.org/10.1016/S0168-9002(03)01368-8}.

\bibitem{CUORE:2016ons}
C.~Alduino, \emph{et~al.}, Measurement of the two-neutrino double-beta decay
  half-life of $^{130}${Te} with the {CUORE-0} experiment. \emph{Eur. Phys. J.
  C} \textbf{77}~(1), 13 (2017),
  \url{https://doi.org/10.1140/epjc/s10052-016-4498-6}.

\bibitem{Wilks:1938dza}
S.~S. Wilks, {The Large-Sample Distribution of the Likelihood Ratio for Testing
  Composite Hypotheses}. \emph{Annals Math. Statist.} \textbf{9}~(1), 60--62
  (1938), \url{https://doi.org/10.1214/aoms/1177732360}.

\bibitem{PhysRevD.57.3873}
G.~J. Feldman, R.~D. Cousins, Unified approach to the classical statistical
  analysis of small signals. \emph{Phys. Rev. D} \textbf{57}, 3873--3889
  (1998), \url{https://link.aps.org/doi/10.1103/PhysRevD.57.3873}.

\bibitem{Fehr200483}
M.~A. Fehr, M.~Rehkamper, A.~N. Halliday, Application of {MC-ICPMS} to the
  precise determination of tellurium isotope compositions in chondrites, iron
  meteorites and sulfides. \emph{Int. J. Mass Spectrometry} \textbf{232},
  83--94 (2004), \url{https://doi.org/10.1016/j.ijms.2003.11.006}.

\end{thebibliography}
\bibliographystyle{sciencemag}

%
%
%
%
%
%


\section*{Acknowledgments}

We dedicate this work to our deceased colleagues, Ettore Fiorini and Stuart J. Freedman, whose contributions and spirit continue to inspire us.

\paragraph*{Funding:} The CUORE Collaboration thanks the directors and staff of the Laboratori Nazionali del Gran Sasso and the technical staff of our laboratories. This work was supported by the Istituto Nazionale di Fisica Nucleare (INFN); the National Science Foundation under Grant Nos. NSF-PHY-0605119, NSF-PHY-0500337, NSF-PHY-0855314, NSF-PHY-0902171, NSF-PHY-0969852, NSF-PHY-1307204, NSF-PHY-1314881, NSF-PHY-1401832, and NSF-PHY-1913374; Johns Hopkins University, Yale University, and University of Pittsburgh. This material is also based upon work supported by the US Department of Energy (DOE) Office of Science under Contract Nos. DE-AC02-05CH11231 and DE-AC52-07NA27344; and by the DOE Office of Science, Office of Nuclear Physics under Contract Nos. DE-FG02-08ER41551, DE-FG03-00ER41138, DE-SC0012654, DE-SC0020423, DE-SC0019316, and DE-SC0011091. This research used resources of the National Energy Research Scientific Computing Center (NERSC). This work makes use of both the DIANA data analysis and APOLLO data acquisition software packages, which were developed by the CUORICINO, CUORE, LUCIFER, and CUPID-0 Collaborations. The authors acknowledge Advanced Research Computing at Virginia Tech for providing computational resources and technical support that have contributed to the results reported within this paper. URL: https://arc.vt.edu/.

\paragraph*{Author contributions:}
All authors contributed to varying degrees in the preparation of the publication, including design, construction, and operations of the detector, collection and analysis of data, and writing and visualization.

\paragraph*{Competing interests:}
There are no competing interests to declare.

\paragraph*{Data and materials availability:} 
The data and software to reproduce the results of this paper are archived on Zenodo~\cite{zenodo}.

\subsection*{Supplementary materials}
Materials and Methods\\
Figs. S1 to S7\\
Tables S1 to S2\\
CUORE Collaboration Author List\\
References \textit{(67-\arabic{enumiv})}\\ 


\newpage


\renewcommand{\thefigure}{S\arabic{figure}}
\renewcommand{\thetable}{S\arabic{table}}
\renewcommand{\theequation}{S\arabic{equation}}
\renewcommand{\thepage}{S\arabic{page}}
\setcounter{figure}{0}
\setcounter{table}{0}
\setcounter{equation}{0}
\setcounter{page}{1} 


\begin{center}
\section*{Supplementary Materials for\\ \scititle}

        CUORE~Collaboration$^{\ast}$\\
	\small$^\ast$Corresponding author: cuore-spokesperson@lngs.infn.it\\
\end{center}

\subsubsection*{This PDF file includes:}
Materials and Methods\\
Figs. S1 to S7\\
Tables S1 to S2\\
CUORE Collaboration Author List

\newpage


\subsection*{Materials and Methods}

\subsubsection*{Denoising}

For this analysis, CUORE uses a nonlinear multivariate noise cancellation algorithm to reduce vibrationally-induced noise that can degrade the detector energy resolution and energy thresholds~\cite{Vetter:2023fas}. The algorithm is designed to take concurrent measurements of nearby environmental noise and noise from the calorimeter readout as inputs and build a transfer function to use as a filter. By applying the filter to the environmental noise, the algorithm predicts the effects of vibrational noise on our calorimeter data stream. This prediction is subtracted from the actual calorimeter data to obtain our denoised data. 

\begin{figure}[h]
    \centering
    \includegraphics[width=0.8\textwidth]{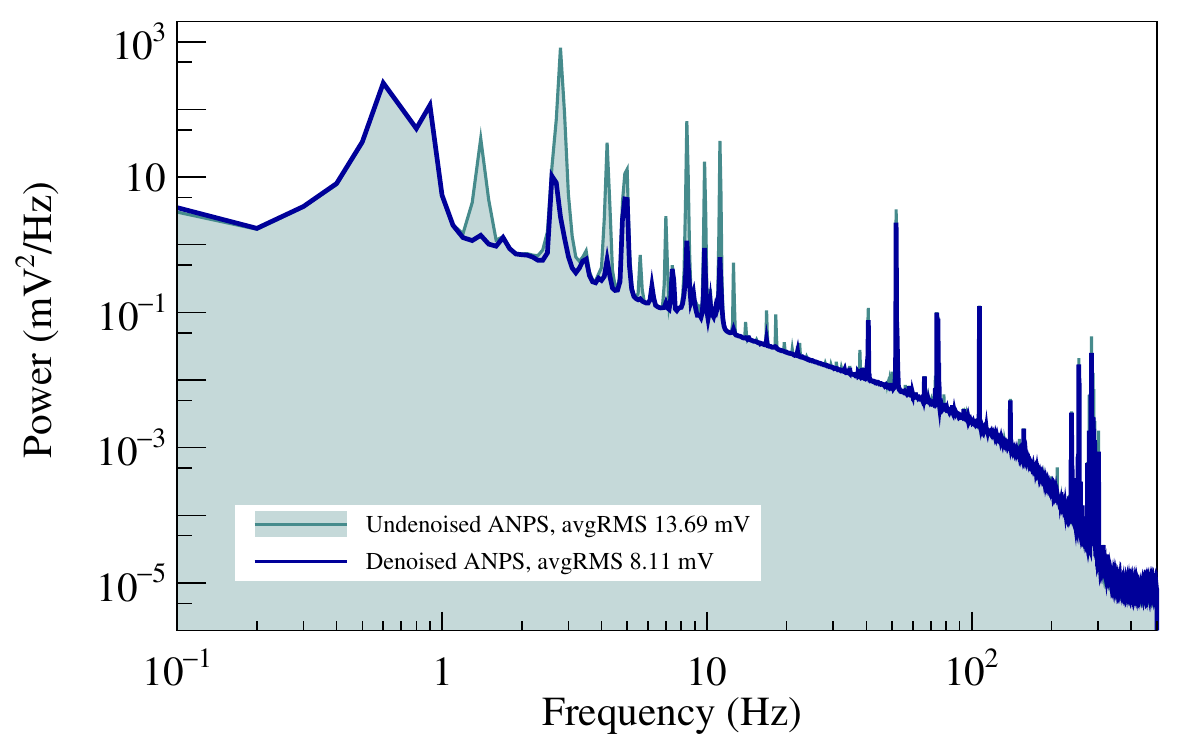} 
    \caption{\textbf{Single calorimeter ANPS.} Average noise power spectra constructed from noise events in denoised (blue) and undenoised (shaded green) physics data of a single dataset.}
    \label{fig:anps}
\end{figure}

We installed microphones around the cryostat in 2017 and expanded the array of auxiliary devices to include three low-frequency accelerometers in 2020. In 2023, we installed a triaxial seismometer and three additional accelerometers. 
For each dataset, we use a subset of these auxiliary devices, each placed around the cryostat, to measure the environmental noise. Noise from the calorimeter readout is taken as noise events---i.e., random samples of the data stream where no pulse is present. In order to correlate the noise with frequency, we take the discrete Fourier transform of each calorimeter noise event ($N$) and each concurrent auxiliary device measurement ($A_i$), where the subscript is used to index the devices, and convolve their spectral densities. For each event, the algorithm uses both the correlated noise between the different auxiliary devices (Eq.~\ref{eq:CA}) and the correlated noise between each auxiliary device and calorimeter (Eq.~\ref{eq:CN}): 
\begin{equation}
\label{eq:CA}
G_{A_i,A_j} \propto \langle A_i^*A_j \rangle
\end{equation}
\begin{equation}
\label{eq:CN}
G_{A_i,N} \propto \langle A_i^*N \rangle\ 
\end{equation}
For each calorimeter, we then calculate the transfer function ($H$) as
\begin{equation}
\label{eq:TF}
H_{A_i,N} = G_{A_i,A_j}^{-1}G_{A_j,N}
\end{equation}
and the denoised data ($D$) as the result of the algorithm on the raw calorimeter measurement ($C$)
\begin{equation}
\label{eq:den}
D = C - H_{A_i,N}^{T}A_i ,
\end{equation}
where the transfer function is only applied to calorimeter measurements $C$ taken during the same time period as the noise measurements used to construct $H$. This time frame is chosen to be $\sim$24~hours, so the algorithm is guaranteed to be responsive to daily variations in noise sources.  
The combination of the aforementioned auxiliary devices is able to measure frequencies between 0.1~Hz -- 1~kHz~\cite{Vetter:2023fas}, 
which largely covers the frequency bandwidth of particle-induced pulses between 0 -- 20~Hz. Figure~\ref{fig:anps} shows the effect of the algorithm on the average noise power spectrum (ANPS) for a single calorimeter, representative of the subgroup of calorimeters highly susceptible to vibrational noise. The reduced amplitudes of the peaks at 1.4~Hz and its harmonics after denoising shows the effectiveness of the algorithm on even the noisiest of calorimeters. In the time since we collected the datasets included in this analysis, to study additional sources of noise, we have further expanded our set of auxiliary devices to include electromagnetic interference (EMI) antennas and upgraded the hardware to improve their quality in the frequency regions of interest. 

\subsubsection*{Event selection}
\begin{figure}[H]
    \centering
    \includegraphics[width=0.99\textwidth]{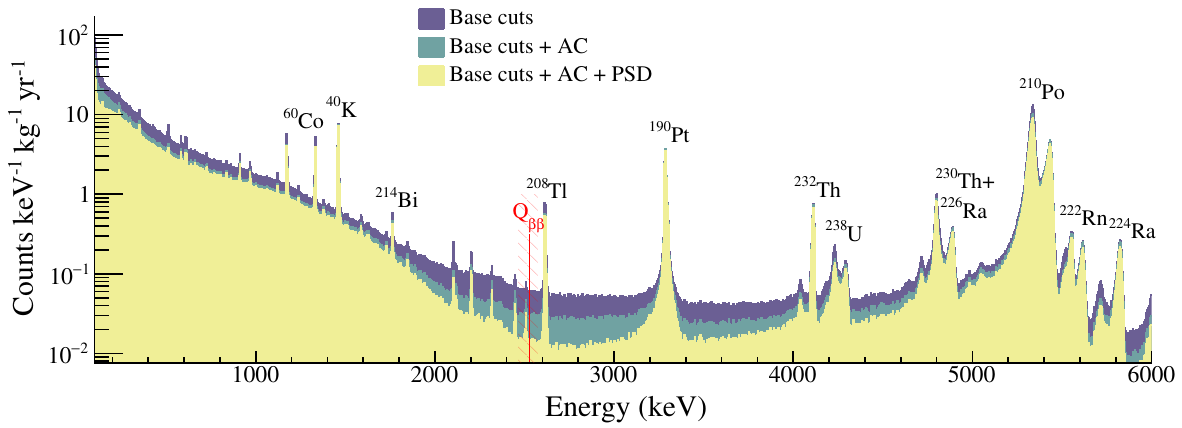} 
    \caption{\textbf{The CUORE energy spectrum after each selection cut.} Physics spectrum for 2039.0~kg$\cdot$yr of TeO$_2$ exposure. We
    show the effects of successively applying the basic quality (base) cuts, the anti-coincidence (AC) cut, and the pulse shape discrimination (PSD) cut. The most prominent $\gamma$ and $\alpha$ peaks in the spectrum are labeled with their radioactive background source. A red line identifies \qbb\ for $^{130}$Te and the hatched region indicates the ROI. }
    \label{fig:spectrum}
\end{figure}

The CUORE energy spectrum after the selection cuts is shown in Fig.~\ref{fig:spectrum}.

\subsubsection*{Detector response evaluation at \ensuremath{\boldsymbol{Q_{\beta\beta}}}}

The detector response function of a monoenergetic peak is modeled empirically after the high-statistics 2615~keV $^{208}$Tl photopeak in calibration data as the superposition of up to three Gaussian distributions that share the same width but have different amplitudes and slightly shifted peak positions (Eq.~\ref{eq:LineshapeFormula}). In the model, the $A_{L,R} < 1$ parameters tune the amplitudes of the left (L) and right (R) sub-peaks with respect to the main peak, whereas the $a_L<1$ and $a_{R}>1$ parameters tune the positions of the means of the sub-peaks with respect to the main peak. Thus, the normalized response function of each calorimeter in each dataset to an energy deposition $E$,
\begin{equation}
    f_{cal}(E;\mu,\sigma,A_L,A_R,a_L,a_R) \doteq \\
    \frac{\mathcal{G}(E;\mu,\sigma) + A_L \mathcal{G}(E;a_L \mu,\sigma) + A_R \mathcal{G}(E;a_R \mu,\sigma)}{ 1 + A_L + A_R}
    \label{eq:LineshapeFormula}
\end{equation}
where $\mathcal{G}(E;\mu,\sigma)$ is a normalized Gaussian distribution with mean $\mu$ and standard deviation $\sigma$, is completely defined by 6 parameters: $\mu$, $\sigma$, $A_{L}$, $A_{R}$, $a_{L}$, and $a_{R}$.

\begin{figure}[H]
    \centering
    \includegraphics[width=\textwidth]{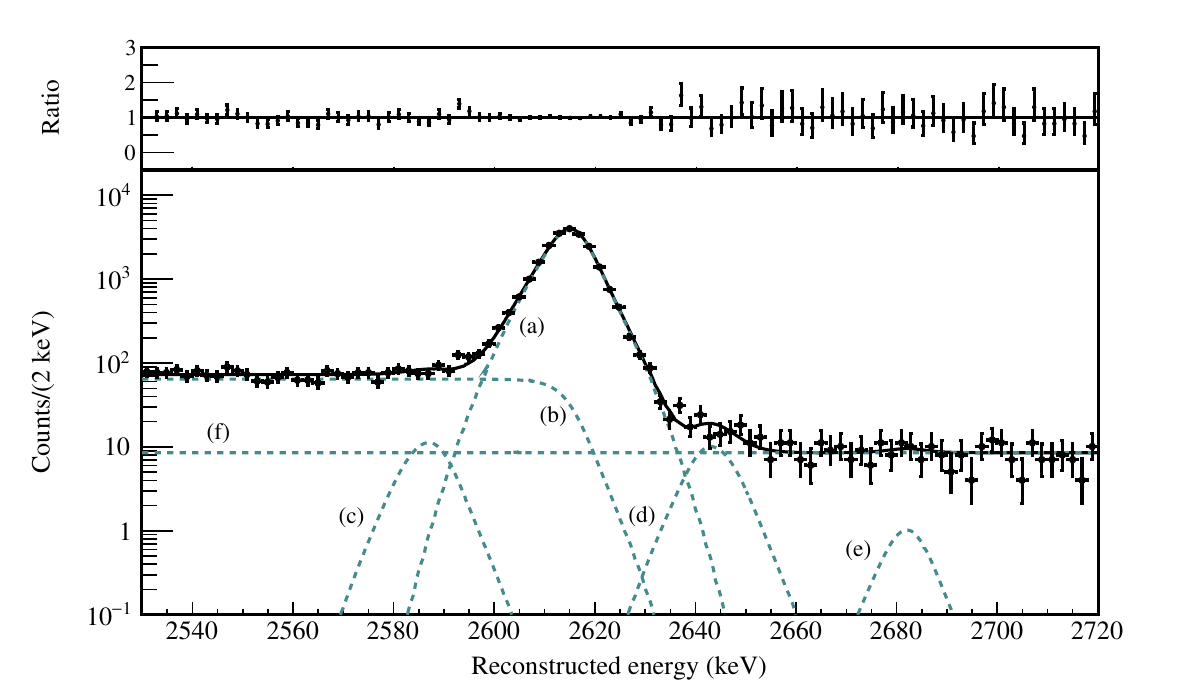} 
    \caption{\textbf{Peak shape extraction.} Bottom: Result of the response function fit on a single tower in calibration data for a single dataset. The solid line is the sum of the best-fit peak shape model over all calorimeters in the tower; the components of this summed best-fit model are shown by the green dashed lines. We identify (a) the multi-Gaussian photopeak, (b) the Compton shoulder, (c) the X-ray escape peak, (d) the X-ray coincidence peak, (e) the coincidence peak at 2687~keV, and (f) a uniform background contribution. A detailed description of the sources of the X-ray peaks is included in the text. Top: Ratio between calibration data and peak shape fit result.}
    \label{fig:ls_calibration}
\end{figure}

We include other components to properly model the energy spectrum near 2615~keV: a multi-Compton scatter contribution; an X-ray escape peak, corresponding to a 2615~keV $\gamma$-ray deposit followed by the escape of one of the 27 -- 31~keV Te X-rays; an X-ray coincidence peak, corresponding to a 2615~keV $\gamma$ ray fully absorbed at the same time as a Te X-ray emitted from a nearby crystal; a coincidence peak at 2687~keV resulting from the simultaneous absorption of 2615~keV and 583~keV $\gamma$ rays from $^{208}$Tl, followed by escape of a 511~keV annihilation $\gamma$ ray; and a uniform background. Due to the lower statistics of these additional components, to constrain their corresponding nuisance parameters, we estimate the response function parameters for each calorimeter in each dataset with a simultaneous, unbinned extended maximum likelihood fit performed on each tower in the energy range [2530,2720]~keV. The result of a typical fit with a breakdown of the fit model is shown in Fig.~\ref{fig:ls_calibration}.

Compared to our previous analysis~\cite{CUORE:2021mvw}, we improved the procedure to fit the lateral sub-peaks by allowing the amplitude and mean ranges to vary after the first fit iteration and omitting any of the lateral sub-peaks in subsequent fit iterations if the relative amplitude was below a set threshold. This means that the photopeak was not constrained to be a triple-Gaussian for every calorimeter, but could also be described with a single- or double-Gaussian. This helped reduce the number of fit parameters as well as the number of iterations required to reach convergence, expediting the whole process.

Since we rely on physics data for our \ndbd\ search, we fit our calibration-based peak-shape model to the most prominent gamma peaks in physics data using the extracted model parameters above and determine their energy resolutions and positions. For each peak-dataset, we perform the fit simultaneously over all active calorimeters in the dataset to ensure adequate statistics. We parametrize the energy resolution in terms of the calibration-based, calorimeter-dependent detector response parameter $\sigma_{cal,i}$ where $i$ identifies the calorimeter. The resolution is modeled as $\sigma_{phys,i,j}=\sqrt{\delta_{phys,i}^2+S_{\sigma,j}\cdot(\sigma_{cal,i}^2-\delta_{cal,i}^2)}$ where $j$ identifies a $\gamma$ peak of a particular energy, $\delta$ represents the baseline resolution measured from noise events from physics data or calibration data, and $S_{\sigma,j}$ is a dataset-dependent global parameter common to all calorimeters. Compared to the model used in our previous analyses, the one described here better captures the actual energy-dependent scaling terms by introducing the intrinsic resolution terms, $\delta$. The peak position is parameterized as $\mu_{phys,j}=S_{\mu,j}\cdot \mu_{nom}$ where $j$ identifies a $\gamma$ peak of a particular energy, $\mu_{nom}$ is the nominal value of the peak energy, and $S_{\mu,j}$ is a dataset-dependent global parameter common to all calorimeters. In order to determine the detector response at \qbb, we model both $S_\sigma$ and the calibration bias, $\Delta\mu$, defined as the difference between the reconstructed and nominal peak energy, as functions of energy, with $S_\sigma$ being described by a phenomenological function that asymptotes to a quadratic polynomial around \qbb\ (paper in preparation), and $\Delta\mu$ likewise being parametrized by a quadratic polynomial. 

\subsubsection*{Statistical analysis for \ndbdbold\ search}

\begin{figure}[H]
    \centering
    \includegraphics[width=0.75\textwidth]{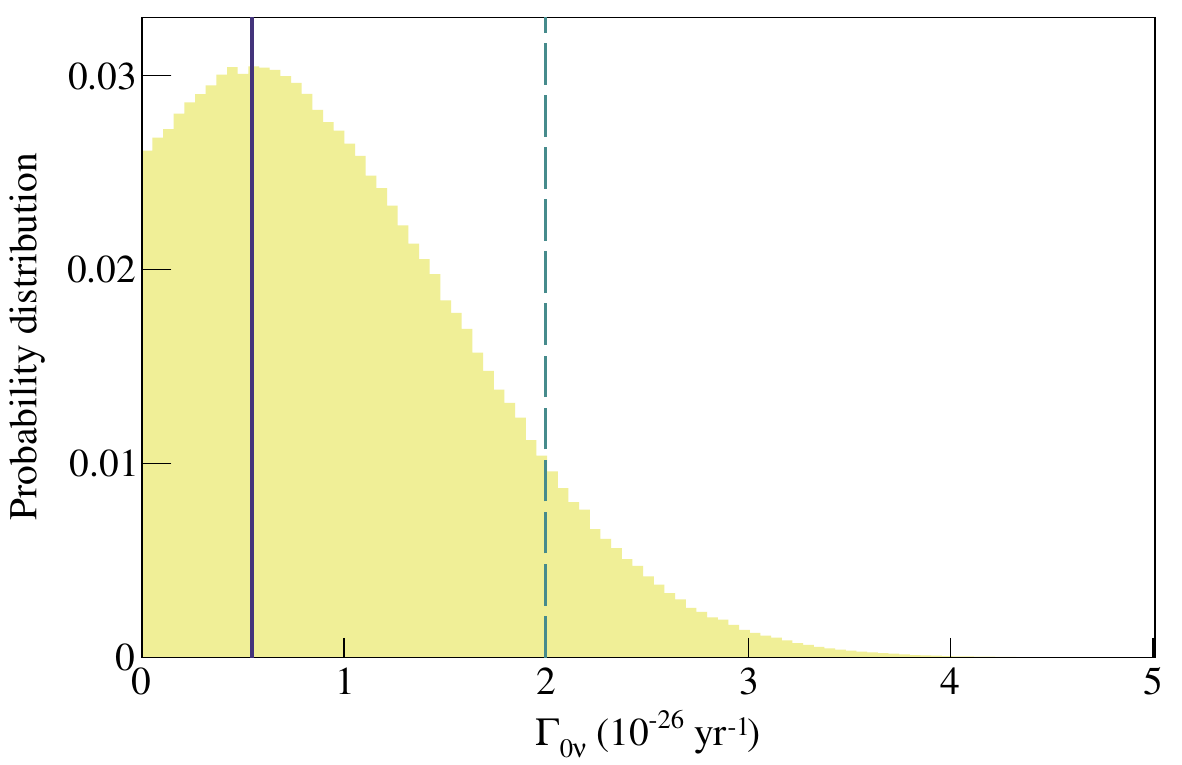} 
    \caption{\textbf{Posterior distribution for the Bayesian fit.} Posterior generated by combining the posteriors to the ROI fits of the two data partitions. We also show the best-fit $\hat\Gamma_{0\nu}$~=~5.5$^{+7.3}_{-5.5}$~(stat.+syst.)~$\times$~10$^{-27}$~yr$^{-1}$~(purple line), which corresponds to the mode of the posterior and the 90\% C.I. upper limit of $ \Gamma_{0\nu}$~$<$~2.0~$\times$~10$^{-26}$~yr$^{-1}$~(green dashed line).} 
    \label{fig:posterior}
\end{figure}

In order to search for an excess of events at \qbb, we perform an unbinned Bayesian fit to the ROI using the Bayesian Analysis Toolkit (\textsc{BAT})~\cite{BATSoftware}. For each calorimeter and dataset, we model the spectrum in the ROI as the superposition of a signal~($s$), a $^{60}$Co background peak~($c$), and a flat background~($b$), and the expectation value of the number of events in the ROI as $\lambda = s + b + c$. The signal~$\big($$^{60}$Co background$\big)$ peak is modeled for each active calorimeter, individually, as $f_{0\nu}$~$\big($$f_\mathrm{Co}$$\big)$, where $f$ has the same form as Eq.~\ref{eq:LineshapeFormula}, but with the mean and variance projected to \qbb\ (2505.7~keV). An additional constraint on the rate of $^{60}$Co is placed at the dataset-level to ensure that the decay follows its known half-life. In order to minimize the computational cost of the search with a large number of parameters, we split the 28 datasets randomly into two approximately equal partitions, carry out the search separately, and combine the posteriors. The combined posterior is shown in Fig.~\ref{fig:posterior}.

\begin{figure}[H]
    \centering
    \includegraphics[width=0.75\textwidth]{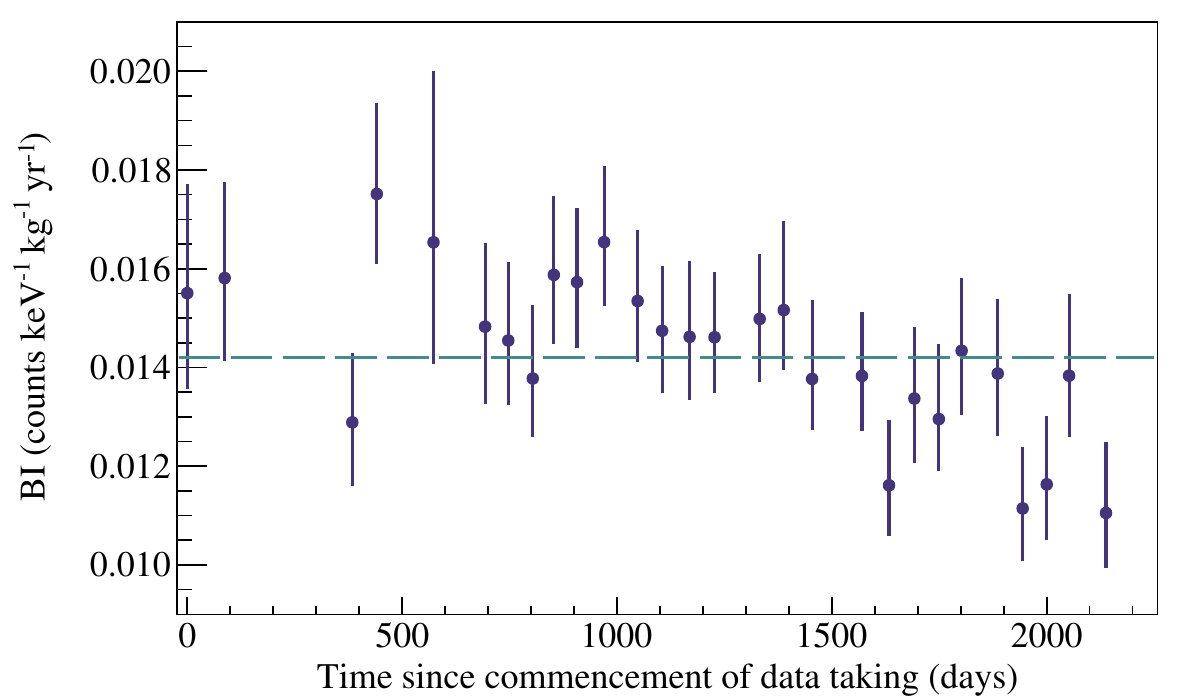} 
    \caption{\textbf{Dataset-dependent background index (BI).} The background index for individual datasets extracted from the background-only fit. Also shown with the dashed green line is the exposure-weighted mean of the background indices over all the datasets.} 
    \label{fig:BI}
\end{figure}

For each dataset (DS) and calorimeter (C), where C is active for the given DS, we use the unbinned extended likelihood: 
\begin{equation}
    \mathcal{L}_{\mathrm{DS,C}} =  \frac{e^{-\lambda} \lambda^{n}}{n!} \prod_{i\in\mathrm{ROI}} \Bigg[ \frac{s}{\lambda} f_{0\nu}(E_{i} | \vec{\theta}_{0\nu}) +  \frac{c}{\lambda} f_\mathrm{Co}(E_{i}| \vec{\theta}_\mathrm{Co}) + \frac{b}{\lambda \Delta E}\Bigg]
\end{equation}
and take the product over all terms as the full likelihood function: 
\begin{equation}
    \mathcal{L} = \prod_{\textrm{DS}}\prod_{\textrm{C}} \mathcal{L}_{\mathrm{DS,C}}
\end{equation}
The total number of measured events in the ROI~($\Delta E$) is denoted by $n$, the reconstructed energy for event $i$ is given by $E_{i}$,
and the model parameters are given by $\vec{\theta}$. By extracting the best-fit value for $s$, we are able to place a limit on the 
\ndbd\ signal rate~($\Gamma_{0\nu}$). The conversion from $s$ to $\Gamma_{0\nu}$ for a given calorimeter accounts for its exposure, efficiency, and the total number of $^{130}$Te atoms contained. We also perform background-only fits of the two partitions without the \ndbd\ signal ($s=0$) to extract the background rate. The background indices for all datasets included in this analysis are shown in Fig.~\ref{fig:BI}.

Systematics are included as nuisance parameters in the \ndbd\ search. The posteriors from the fits characterizing the detector performance-related systematics (i.e., efficiencies, energy bias, and resolution scaling) are included as priors in the \ndbd\ analysis. Efficiencies fall into three broad categories: containment efficiency, reconstruction efficiency, and event selection efficiency. Containment efficiency---characterizing the percentage of \ndbd\ events that deposit their total energy within a crystal---is evaluated using \textsc{Geant4}~\cite{GEANT4:2002zbu} Monte-Carlo simulations~\cite{CUORE:2016ons}. Reconstruction efficiencies encompass all relevant efficiencies at the single pulse level, including the ability to trigger, reconstruct correct energies, and contain a single pulse within a window. Configurable injected heater pulses are good proxies for physical pulses and are used to evaluate the reconstruction efficiencies for each dataset-calorimeter. Selection efficiencies include anti-coincidence and pulse shape discrimination (PSD) efficiencies which express the probability of retaining signal-like events after applying the single-crystal cut and PSD cut, respectively. These are evaluated numerically at the dataset level by using known radioactive decay peaks~\cite{CUORE:2021mvw}. 

A list of all the relevant parameters for the current analysis is provided in Table~\ref{tab:Efficiencies}. We extract the analysis efficiencies using a Bayesian counting analysis and combine them to obtain a probability density function for the total analysis efficiency, which is passed as a prior to the \ndbd\ fit. For the detector response parameters, we fit the dataset-level scaling factors for the resolution and calibration bias, $S_{\sigma}(E)$ and $\Delta\mu(E)$, respectively, with a Bayesian approach using \textsc{BAT} to obtain posterior probability distributions, which are treated as priors in the \ndbd\ fit. Finally, following a conservative approach, we choose a non-informative, uniform prior for $\Gamma_{0 \nu}$. A summary of the systematics as well as their sources and priors is shown in Table~\ref{tab:systematics}.

\begin{figure}[H]
    \centering
    \includegraphics[width=0.7\textwidth]{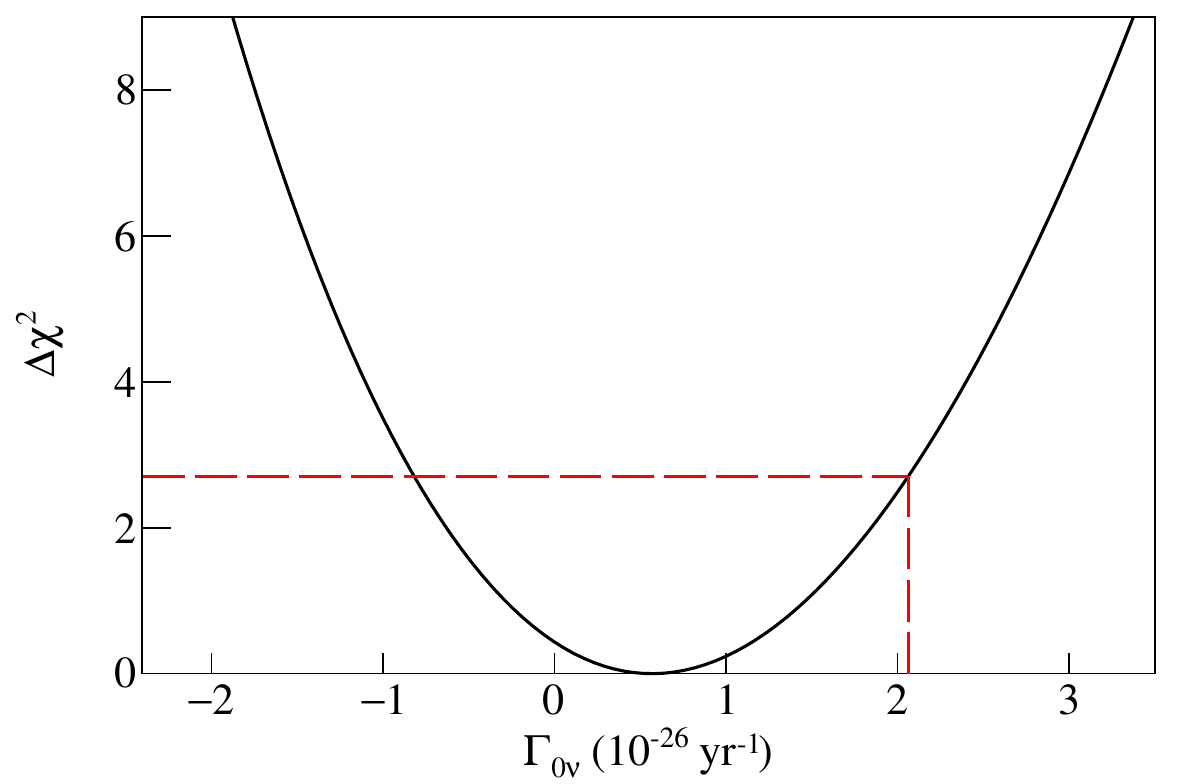} \caption{\textbf{\ensuremath{\boldsymbol{\Delta \chi^{2}}} values from the frequentist fit.} Profile of the combined $\Delta \chi^{2}$, approximated as $-2 \log(\mathcal{L})$, extracted with the frequentist fit. The dashed red line indicates the 90\% C.L.}
    \label{fig:posteriorNLL}
\end{figure}

In addition to the Bayesian analysis, we also perform a frequentist fit using the profile likelihood ratio test statistic~\cite{Algeri:2019lah}. For the frequentist analysis, we use the same likelihood function used for the Bayesian analysis and combine it with the Lagrange multipliers for the efficiency and detector response parameters. We first compute $\Gamma^{M}_{0\nu}$ corresponding to the global maximum of $\mathcal{L}$ while allowing all the parameters to float. We then profile over a range of fixed $\Gamma$s around $\Gamma^{M}_{0\nu}$ and generate a distribution of $\Delta \chi^{2} = -2\ln{\frac{\mathcal{L}(\Gamma)}{\mathcal{L}(\Gamma^{M}_{0\nu})}}$ as shown in Fig.~\ref{fig:posteriorNLL}. Assuming Wilks' theorem~\cite{Wilks:1938dza}, we extract the confidence interval in $\Gamma$ at $\Delta \chi^{2} =$ 2.706 which corresponds to the 90\% confidence level (C.L.). We check the validity of Wilks' theorem in the context of our application using ensembles of toy experiments. We find consistent statistical coverage and limits between confidence intervals derived using Wilks' theorem, and those built from toy experiments following a Feldman-Cousins construction~\cite{PhysRevD.57.3873}.

\begin{figure}[H]
    \centering
    \includegraphics[width=0.65\textwidth]{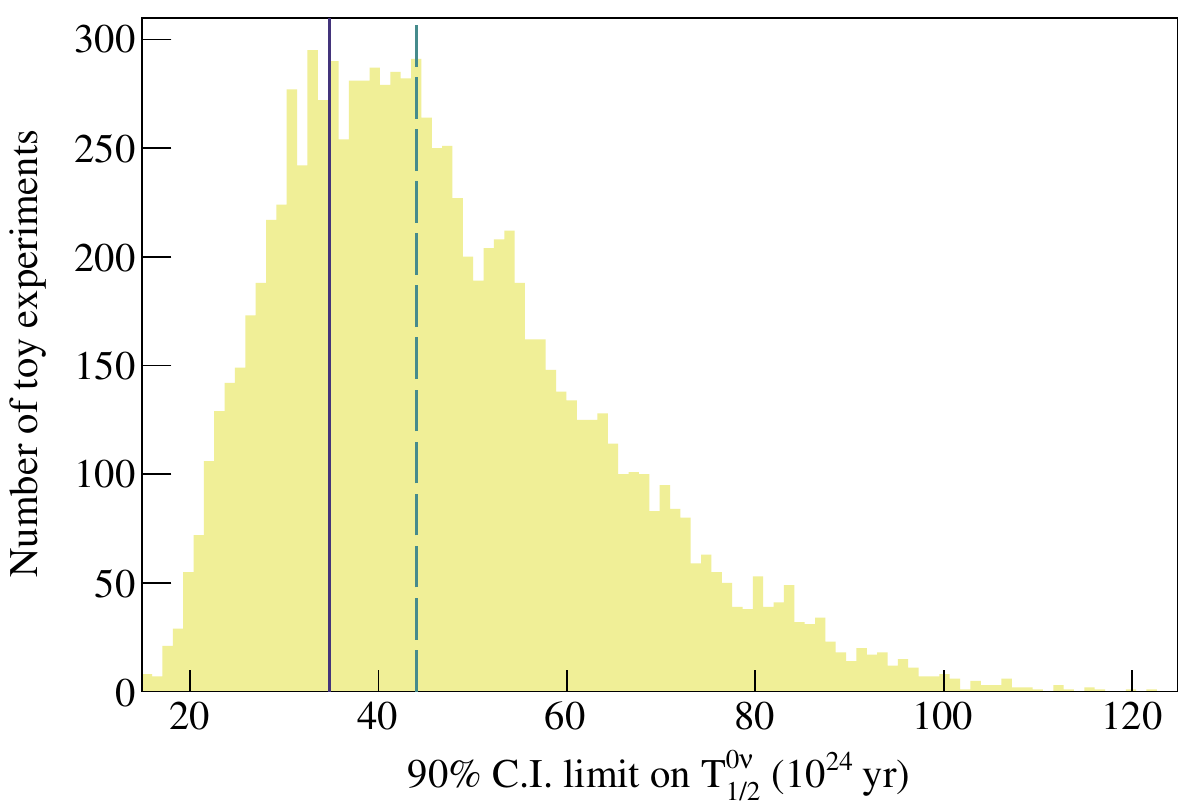} 
    \caption{\textbf{Bayesian exclusion sensitivity.} The distribution of the \ndbd\ half-life limits generated using 10$^4$ background-only toy Monte Carlo simulations and fit to a signal+background model with the Bayesian approach. Also shown is the \ndbd\ half-life limit~(purple line) set by CUORE for comparison with the median~(dashed green line) of the sensitivity.}    
    \label{fig:excl_sens}
\end{figure}

Employing the same likelihood function, we also extract the experimental sensitivity using 10$^4$ background-only toy Monte Carlo experiments. A search for the presence of a peak at \qbb\ is then performed using Bayesian~(frequentist) analysis under the signal+background hypothesis and a 90\% C.I.~(C.L.) distribution on the half-life of \ndbd\ is generated as shown, for the Bayesian approach, in Fig.~\ref{fig:excl_sens}. Given the median of the distribution---called the exclusion sensitivity---at 4.4~$\times$~10$^{25}$~yr~(4.7 $\times$ 10$^{25}$~yr), the probability of obtaining a more stringent limit is given by 74\%~(73\%) using the Bayesian~(frequentist) analysis. We also generate \ndbd\ signal-inclusive toy Monte Carlo experiments over the $\Gamma_{0\nu}$ range of 1 $\times$ 10$^{-26}$ -- 1 $\times$ 10$^{-25}$ yr$^{-1}$---corresponding to the $T^{0\nu}_{1/2}$ range of 6.9 $\times$ 10$^{24}$ -- 6.9 $\times$ 10$^{25}$ yr--- and perform a frequentist \ndbd-inclusive fit to extract the median \ndbd\ discovery sensitivity of $T^{0\nu}_{1/2} > $ 2.6 $\times$ 10$^{25}$ yr (3$\sigma$ C.L.)---a value 1.8 times weaker than the frequentist exclusion sensitivity.

\newpage


\begin{table}[H]
  \centering
  \caption{
  \textbf{Summary of relevant parameters for the \ndbdbold\ analysis.} 
    The resolution and efficiencies are exposure-weighted mean values.}\label{tab:Efficiencies}
  \begin{tabular}{ll}
    \toprule
    \midrule
    Number of datasets & 28 \\
    TeO$_2$ exposure & 2039.0~kg$\cdot$yr\vspace{1pt}\\
    $^{130}$Te exposure & 567.0~kg$\cdot$yr\vspace{1pt}\\
    \hline\vspace{-20pt} \\
    FWHM at 2615 keV in calibration data & 7.540(24)~keV\\
    FWHM at \qbb\ from physics data & 7.310(24)~keV\vspace{1pt}\\ \hline\vspace{-20pt} \\
    Total analysis efficiency (data) & 93.4(18)\%\\ 
    \quad Reconstruction efficiency & 95.624(16)\%\\ 
    \quad Anti-coincidence efficiency & 99.80(5)\%\\
    \quad PSD efficiency & 97.9(18)\%\\ 
    Containment efficiency (Monte Carlo) & 88.35(9)\%~\cite{Alduino:2016vjd} \\
    \midrule
    \bottomrule
  \end{tabular}   
\end{table}

\begin{table}[H]
  \centering
  \caption{\textbf{Summary of fit parameters.} 
  The list of fit parameters including systematics with their priors. We distinguish whether the fit parameters are allowed to vary by dataset~(dataset) or are constrained to vary identically across all the datasets in the fit~(global).
  Since the \ndbd\ rate, $^{130}$Te \qbb, and isotopic abundance are predicted to remain constant with time, they are regarded as global parameters. Because detector performance parameters could have dataset dependence and the background index and the decay rate of the $^{60}$Co sum peak could have time dependence, they are regarded as dataset parameters.
  }\label{tab:systematics}
  \begin{tabular}{llll}
    \toprule
    \midrule
    Parameter & Prior & Dependence & Source \\
    \midrule
    $\Gamma_{0\nu}$ & Uniform & Global & - \\
    $\Gamma_{\mathrm{Co}}$ & Uniform & Dataset & - \\
    Background index & Uniform & Dataset & - \\
    \midrule
    $^{130}$Te \qbb & Gaussian & Global & Literature values~\cite{Rahaman:2011zz} \\
    Isotopic abundance & Gaussian & Global & Literature values~\cite{Fehr200483} \\
    Containment efficiency & Gaussian & Global & Monte-Carlo simulation \\
    Total analysis efficiency & Multivariate & Dataset & Fit Posteriors \\
    Energy bias and resolution & Multivariate & Dataset & Fit Posteriors \\
    \midrule
    \bottomrule
  \end{tabular}   
\end{table}

\newpage

\subsection*{CUORE Collaboration Author List}

\noindent
D.~Q.~Adams$^{1}$, C.~Alduino$^{1}$, K.~Alfonso$^{2}$, F.~T.~Avignone~III$^{1}$, O.~Azzolini$^{3}$, G.~Bari$^{4}$, F.~Bellini$^{5,6}$, G.~Benato$^{7,8}$, M.~Beretta$^{9}$, M.~Biassoni$^{10}$
% ...

\vspace{8mm}

\noindent
{\small
$^{1}$ Department of Physics and Astronomy, University of South Carolina, Columbia, SC 29208, USA \\
$^{2}$ Center for Neutrino Physics, Virginia Polytechnic Institute and State University, Blacksburg, Virginia 24061, USA \\
$^{3}$ INFN -- Laboratori Nazionali di Legnaro, Legnaro (Padova) I-35020, Italy \\
$^{4}$ INFN -- Sezione di Bologna, Bologna I-40127, Italy \\
$^{5}$ Dipartimento di Fisica, Sapienza Universit\`{a} di Roma, Roma I-00185, Italy \\
$^{6}$ INFN -- Sezione di Roma, Roma I-00185, Italy \\
$^{7}$ Gran Sasso Science Institute, L'Aquila I-67100, Italy \\
$^{8}$ INFN -- Laboratori Nazionali del Gran Sasso, Assergi (L'Aquila) I-67100, Italy \\
$^{9}$ Department of Physics, University of California, Berkeley, CA 94720, USA \\
$^{10}$ INFN -- Sezione di Milano Bicocca, Milano I-20126, Italy \\
% ...
}



\end{document}